\documentclass[lettersize,journal]{IEEEtran}
\usepackage{amsmath,amsfonts}
\usepackage{algorithmic}
\usepackage{algorithm}
\usepackage{array}
\usepackage[caption=false,font=normalsize,labelfont=sf,textfont=sf]{subfig}
\usepackage{textcomp}
\usepackage{stfloats}
\usepackage{url}
\usepackage{verbatim}
\usepackage{graphicx}
\usepackage{cite}
\usepackage{amsfonts,amssymb} 

\usepackage{bbding}
\usepackage{booktabs} 
\usepackage{multirow}

\usepackage{color}

\newcommand{\mybluehl}[1]{\textcolor{black}{#1}}

\begin{document}

\title{Soundscape Captioning using Sound Affective Quality Network and Large Language Model}


\author{Yuanbo Hou, Qiaoqiao Ren, Andrew Mitchell, Wenwu Wang, Jian Kang, Tony Belpaeme, Dick Botteldooren

\thanks{ 



\vspace{-4mm}



This research received funding from Flemish Government under the “Onderzoeksprogramma Artificiële Intelligentie (AI) Vlaanderen” programme.


Yuanbo Hou is with the Machine Learning Group, University of Oxford, UK. Corresponding email: Yuanbo.Hou@eng.ox.ac.uk

Qiaoqiao Ren and Tony Belpaeme are with the AIRO-IDLab, Ghent University-Imec, Belgium. 

Wenwu Wang is with the CVSSP, University of Surrey, Guildford, UK.

Andrew Mitchell and Jian Kang are with University College London, UK.

Dick Botteldooren is with the WAVES Group, Ghent University, Belgium.
} 
}



\maketitle

\begin{abstract}  
We live in a rich and varied acoustic world, which is experienced by individuals or communities as a \emph{soundscape}. 
Computational auditory scene analysis, disentangling acoustic scenes by detecting and classifying events, focuses on objective attributes of sounds, such as their category and temporal characteristics, ignoring their effects on people, such as the emotions they evoke within a context. 
To fill this gap, we propose the \mybluehl{affective soundscape captioning (ASSC)} task, which enables automated soundscape analysis, thus avoiding labour-intensive subjective ratings and surveys in conventional methods. 
With soundscape captioning, context-aware descriptions are generated for soundscape by capturing the acoustic scenes (ASs), audio events (AEs) information, and the corresponding human affective qualities (AQs).
To this end, we propose an automatic soundscape captioner (SoundSCaper) system composed of an acoustic model, i.e. SoundAQnet, and a large language model (LLM). 
SoundAQnet simultaneously models multi-scale information about ASs, AEs, and perceived AQs, while the LLM describes the soundscape with captions by parsing the information captured with SoundAQnet.
\mybluehl{
SoundSCaper is assessed by two juries of 32 people. In expert evaluation, the average score of SoundSCaper-generated captions is slightly lower than that of two soundscape experts on the evaluation set D1 and the external mixed dataset D2, but not statistically significant. 
In layperson evaluation, SoundSCaper outperforms soundscape experts in several metrics on datasets D1 and D2. 
In addition to human evaluation, compared to other automated audio captioning (AAC) systems with and without LLM, SoundSCaper performs better on the ASSC task in several natural language processing (NLP) based metrics. 
Overall, SoundSCaper performs well in human subjective evaluation and various objective captioning metrics, and the generated captions are comparable to those annotated by soundscape experts.
} 
The model, source code, LLM scripts, human assessment data, instructions, and evaluation statistics are all publicly available.
\end{abstract}

\vspace{-1mm}

\vspace{-2mm}
\section{Introduction}\label{Introduction}

\IEEEPARstart{S}oundscape \mybluehl{plays a vital role in shaping our daily experience, affecting various aspects including mood, behaviours, and overall well-being \cite{erfanian2021psychological}. Understanding how people perceive soundscape has become important in urban design, environmental psychology, and interactive technology \cite{erfanian2019psychophysiological}}. 
The international standard organization (ISO) 12913-1:2014 \cite{ISO_soundscape} defines soundscape as: ``\textit{the acoustic environment as perceived or experienced and/or understood by a person or people, in context}", \mybluehl{which emphasizes the interaction between a person or people and the acoustic environment.}
\mybluehl{Thus, a soundscape is not only about the audio events (AEs) present in the acoustic scene (AS), but also the emotions and mood evoked by the acoustic environment and events therein.}

\begin{figure}[b]
	\setlength{\abovecaptionskip}{-1mm}  
    \setlength{\belowcaptionskip}{-5cm}   
      \vspace{-6mm}
	\centerline{\includegraphics[width = 0.3 \textwidth]{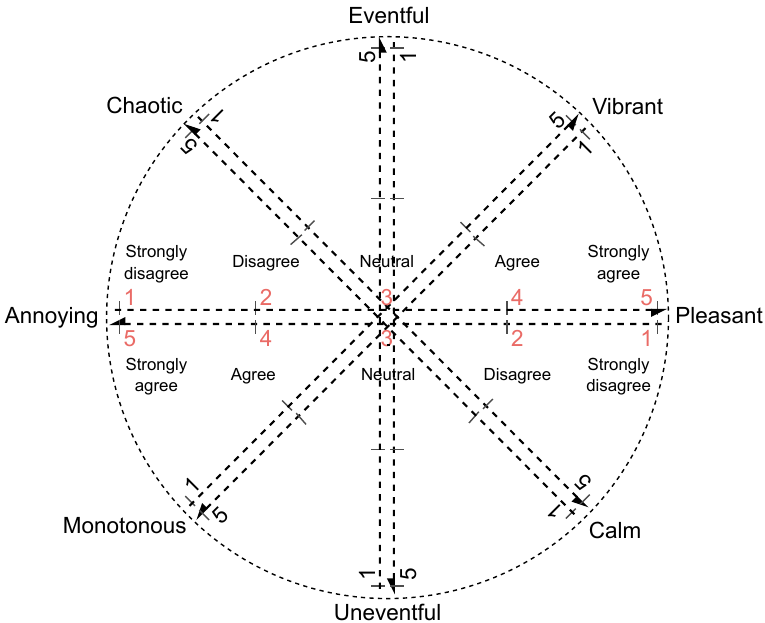}}
 \caption{The 8-dimensional (8D) affective qualities (AQs) in the soundscape circumplex model (SCM) \cite{Axelsson2010principal} recommended by ISO/TS 12913-3:2019 \cite{ISO_soundscape_PAQ}.}
	\label{SCM}
\end{figure}

To describe the affect of soundscape, ISO/TS 12913-3:2019 \cite{ISO_soundscape_PAQ} recommends using the soundscape circumplex model (SCM) \cite{Axelsson2010principal}, a framework inspired by the affect theory of emotions \cite{russell1980circumplex}. 
In Fig. \ref{SCM}, the SCM is scored on eight 5-point Likert scales 
to describe the perceptual attributes of soundscapes.
Some prior studies \cite{hou23_interspeech}\cite{icassp2024} explore the relationships between AEs and annoyance, which is one of the 8 attributes of perceived affective quality (PAQ) in SCM. 
These perceptions are shaped by sound characteristics, contextual cues, and prior experience or common knowledge specific to culture. 
\mybluehl{In addition, the affect of soundscape is highly related to the perception of AS as a whole. Therefore, to describe a soundscape, it is crucial to exploit both the physical information about the acoustic environment that can be estimated using techniques such as acoustic scene classification (ASC) and audio event classification (AEC), and the perceptual information, such as human-perceived affective quality (AQ) of the soundscape.}

To enable machines to understand acoustic environments, computational analysis of audio scenes and events \cite{casa} has been studied extensively, e.g., by the detection and classification of acoustic scenes and events (DCASE) community \cite{dcase2017}. 
\mybluehl{This has led to significant advancements in the recognition of ASs and AEs, resulting in various methods, from machine learning methods to deep learning methods, such as AE detection methods \cite{dcase2017}\cite{icassp2019}\cite{pann} based on frame-level strong labels or clip-level weak labels, and ASC methods \cite{dcase2019_t1a}\cite{ergl_spl}\cite{taslp_cSEM}.}  
\mybluehl{More recently, natural language has been used to describe audio content, including ASs and AEs, leading to an emerging area called automated audio captioning (AAC) \cite{audio_caption} \cite{Clotho_dataset}.}


However, the DCASE-related works focus mainly on the objective attributes of sounds, while ignoring the effects these sounds have on people, such as the different emotions they evoke. 
\mybluehl{This results in the relations between various AEs and 8-dimensional (8D) affective qualities (AQs) in PAQ remaining unexplored, let alone the relations between AS, AEs, and affective responses to 8D AQs.}
For example, the AAC systems often describe AEs or AS-related information, such as ``\textit{birds are of chirping the chirping and various chirping}" in the AAC task in DCASE 2020 \cite{Clotho_dataset}, but they have not explored the listener's response to the audio along the affective dimension, i.e., whether hearing birds chirping brings pleasure or annoyance to the listener. 
\mybluehl{Despite the substantial progress in AAC for the acoustic environment, little attention has been paid to the affective information carried by AEs in a soundscape or the soundscape as a whole. More specifically, there is a significant research gap in captioning that links the acoustic environments (ASs and AEs) and the human-perceived AQs of soundscapes.}

\mybluehl{To fill this research gap, we propose a new task, called \textit{affective soundscape captioning} (ASSC), where a soundscape is described using context-aware texts, detailing AS, AE, and emotion-related AQ. 
This enables the exploration of affective information from anthropocentric soundscapes, as defined in ISO 12913-1:2014 \cite{ISO_soundscape}. 
Inspired by recent advancements in large language models (LLMs) \cite{openai_chatgpt}, we propose an LLM-based soundscape caption system (SoundSCaper) for the ASSC task by integrating coarse-grained ASs, fine-grained AEs, and human-perceived AQs. 
SoundSCaper integrates the rich prior knowledge in LLMs like generative pre-trained transformer (GPT) models \cite{GPT4}, enabling context-sensitive and affectively meaningful descriptions of soundscapes.
}

\mybluehl{
To provide precise acoustic inputs to the LLM, we propose a sound affective quality network (SoundAQnet) to simultaneously model AS and AE in acoustic environments, and their corresponding AQ. SoundAQnet combines the Mel-spectrogram features \cite{pann} commonly used in ASC and AEC tasks with loudness (Zwicker loudness, defined in ISO 532-1 \cite{ISO_loudness}) to capture AS, AE, and AQ in the soundscape. Loudness, which measures the subjective impression of human perception of sound, has important implications for AQ modelling, as the perceived loudness of a sound can change its contribution to affect, from gentle background to extremely disturbing sounds. 
Integrating the output from SoundAQnet with GPT allows SoundSCaper to generate natural language descriptions, thereby linking objective acoustic indicators (i.e., AS and AE) with human-centred perceptual indicators (i.e., PAQ). 
To our knowledge, we are the first to build models that simultaneously characterize AS, AE, and AQ in a soundscape and to describe the soundscape with affective captions using an LLM based on the three-view information (i.e., AS, AE, and AQ).}


This paper strives to advance machine listening by linking it with affective computing and contextual interpretation, thus going beyond conventional recognition and classification of sounds. Our work offers the potential to enable machines to have a comprehensive and emotionally attuned perception of auditory scenes and events.
The contributions of this paper are as follows:
1) We propose the ASSC task, where a soundscape is described in free texts from the perspectives of AS, AE, and AQ, thus bridging the gap between audio captions and the human-perceived AQs of sounds. 
2) We propose SoundAQnet to simultaneously model the coarse-grained AS and fine-grained AE, as well as human-perceived AQ.
3) \mybluehl{We utilize the rich knowledge embedded in LLM about expected sounds in various scenes to develop the automatic SoundSCaper system, which translates SoundAQnet’s predictions into human-understandable natural language captions;  to this end, careful soundscape-focused prompt engineering is introduced.
This transforms soundscape descriptions from limited numerical values into comprehensible free text rich in acoustic context and human AQ perception.}
4) To measure the quality of the soundscape captions, we introduce the Transparent Human Benchmark for Soundscapes (THumBS) as a metric and evaluate the performance of SoundSCaper on the test set and the mixed external dataset.



Next, Section \ref{Related work} discusses related work. 
Section \ref{Soundscape_caption} introduces \mybluehl{the proposed ASSC task.}
Section \ref{LLM_SSD_framework} proposes the SoundSCaper based on SoundAQnet and LLM. 
Section \ref{AM_section_setup} presents SoundAQnet experiments.
Section \ref{section_human_Evaluation} presents \mybluehl{human evaluation experiments on SoundSCaper by expert and layperson groups, compares the results of SoundSCaper and AAC systems on the ASSC task}, and discusses their characteristics. Section \ref{section_conclusion} concludes.
We have released the code and models, human assessment data, and human evaluation statistics to the 
\textbf{\textit{homepage}} (\textcolor{blue}{\underline{https://github.com/Yuanbo2020/SoundSCaper}}).


\vspace{-1.5mm}
\section{Related Work}\label{Related work}


This section reviews related work on soundscape captioning.

\vspace{-2.2mm}
\subsection{Audio Captioning}



\mybluehl{
Audio captioning \cite{audio_caption}\cite{Clotho_dataset} resembles soundscape captioning when it comes to disentangling the auditory scene into separate sounds. Various automated audio captioning (AAC) systems 
\cite{koepke2022audio, labb2024conette, xiao2022local} 
aim to describe AEs and physical properties of acoustic environments using text. ConvNeXt-Trans \cite{labb2024conette}, the baseline of DCASE 2024 challenge \cite{Clotho_dataset}, excels at the AAC task. The audio encoder of ConvNeXt-Trans is ConvNeXt, and its decoder consists of a Transformer decoder with a structure similar to GPT. P-LocalAFT \cite{xiao2022local} allows local information to be captured while retaining global information, which captures AEs of different durations for precise captions. These AAC systems focus on AEs and ASs, as a result, they can describe the objective information of acoustic environments. However, they fail to capture emotions and moods perceived by humans, as the datasets used for model training lack AQ labels.
} 

\mybluehl{
In addition to AAC systems without LLM, audio-LLM-based GAMA \cite{ghosh2024gama} and Qwen-Audio \cite{chu2023qwen} perform well on multiple audio captioning datasets. GAMA uses an acoustic model to extract the information and then feeds it to an LLM to generate captions. GAMA's framework is similar to SoundSCaper. 
Although audio-LLMs \cite{ghosh2024gama}\cite{chu2023qwen} have some audio perception capability, such perception focuses on the objective content of audio and serves audio understanding and reasoning; i.e., it is not the affective perception in soundscapes, let alone the perception of 8D AQs in ISO/TS 12913-3:2019.}

\vspace{-2.2mm}
\subsection{Affective Computing}
\mybluehl{In soundscape affective computing, arousal-valence dimensional models (AVDM) \cite{erfanian2019psychophysiological} are often used to capture human emotional responses. However, AVDM has limited ability to distinguish different affective perceptions, making it difficult to model complex emotion states \cite{wang2024assessing}. Based on the pleasantness-eventfulness framework, the performance of multiple acoustic features on urban soundscapes is analyzed, and the results \cite{lunden2016urban} show that Mel-based features predict the pleasantness and eventfulness of soundscapes well. And Gammatone cepstral coefficients \cite{di2024exploring} have been shown to be feasible in assessing the valence and arousal of sounds. However, the study \cite{di2024exploring} is based on synthetic rather than real datasets. This paper models human affective responses based on a multi-user annotated real soundscape dataset according to 8D AQs in ISO standards.}





\vspace{-3.5mm}
\subsection{Soundscape Analysis}
\mybluehl{Soundscape analysis can be roughly categorized into acoustic environment-oriented \cite{casa, dcase2017, icassp2019, pann} and affective perception-oriented \cite{erfanian2021psychological}\cite{Axelsson2010principal}\cite{soundscape_IADS,ISD,AI_soundscape}. The former 
focuses on the \textit{objective} acoustic environment understanding, e.g., ASC and AEC.  
The latter focuses on affective perceptions of soundscapes to characterize emotional states.
The international affective digital sound (IADS) \cite{soundscape_IADS} dataset explores discrete emotional categories elicited by 167 individual stimuli, and the international soundscape database (ISD) \cite{ISD} explores the relationship between annoyance and 24 categories of AEs in daily life.  
In addition, ARAUS \cite{araus} studies the unique subjective perceptual responses of 605 participants to 25440 augmented soundscapes presented as audiovisual stimuli.
}

\mybluehl{To the best of our knowledge, there are no studies on automatic descriptions of soundscapes, especially AASC.
The successful application of the proposed AASC will promote automated soundscape analysis, improve urban soundscape planning}, and improve environmental awareness of visually and hearing-impaired people \cite{coelho2016approaches}\cite{skagerstrand2014sounds}.

\begin{figure}[b] 
	\setlength{\abovecaptionskip}{-1mm}  
    \setlength{\belowcaptionskip}{0cm} 
    \vspace{-4mm}
	\centerline{\includegraphics[width = 0.48 \textwidth]{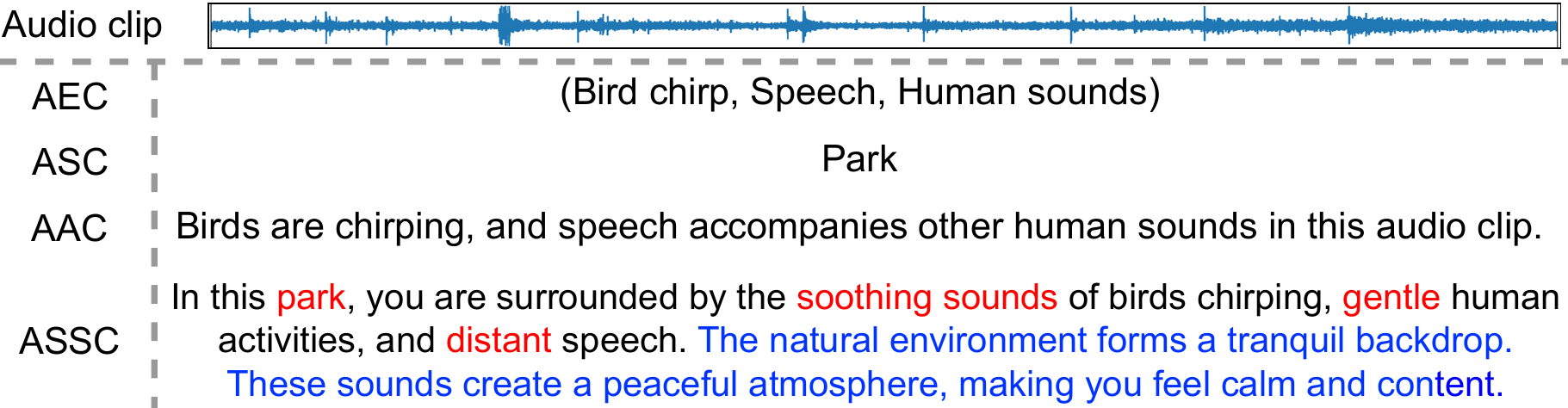}} 
 \caption{Results of tasks: coloured texts show differences between automated audio captioning (AAC) and affective soundscape captioning (ASSC) tasks. Blue texts indicate \textit{human-perceived AQ}-related descriptions unique to ASSC.}
	\label{task_differences}
\end{figure}

\vspace{-2mm}
\section{Affective Soundscape Captioning}
\label{Soundscape_caption}

\mybluehl{This section presents problem formulation and task definition for ASSC, and the SoundScaper system for AASC.}

\vspace{-3mm}
\subsection{Problem Formulation and Task Definition}

\mybluehl{Affective soundscape captioning (ASSC) describes the holistic experience and understanding of an acoustic environment.} 
Identifying and recognizing AEs and assigning meaning to them is one of the cornerstones of this personal and individual experience. Because of this, soundscape evaluation questionnaires (e.g., ISO/TS 12913-2 \cite{ISO_soundscape_data}) inquire about the classes of sounds people hear. Thus, \mybluehl{ASSC should include a description of the relevant categories of audible sounds.}

The holistic evaluation is also affected by psychoacoustics-related loudness \cite{ISO_loudness} and the context of sounds. 
\mybluehl{
Therefore, from the perspective of holistic experience and understanding, ASSC should consider the categories of sounds, the context in which they occur, and the AQ they may evoke in humans.}
Fig.~\ref{task_differences} contrasts ASSC to related tasks. The AEC task aims to tag the types of AEs of audio clips with semantic labels; the ASC task identifies the environment category where the sound is recorded, i.e., its context.
\mybluehl{
The AAC task converts audio content, mainly AEs, into text. 
ASSC starts from the perspective of soundscape with human perception rather than just AEs, adds AS information, and focuses on emotional impact. 
AASC adds affective attributes to the textual description and suggests the human affect evoked by the acoustic environment.
}

\vspace{-3mm}
\subsection{Overview of the Proposed SoundScaper System}

\begin{figure}[b] 
	\setlength{\abovecaptionskip}{-2mm}  
    \setlength{\belowcaptionskip}{-1mm}   
    \vspace{-6mm}
	\centerline{\includegraphics[width = 0.48 \textwidth]{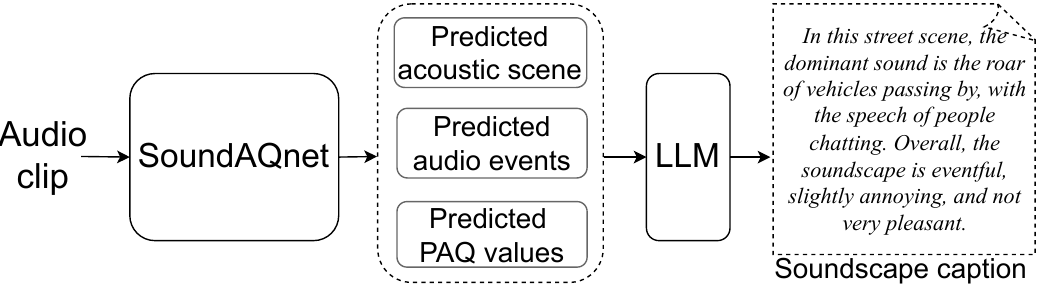}}
	\caption{The proposed automatic soundscape captioner (SoundSCaper) system.}
	\label{LLM_SSD_overall}
\end{figure}

The proposed automatic soundscape captioner, SoundSCaper, as shown in Fig. \ref{LLM_SSD_overall}, consists of two parts: the acoustic model ($am(\cdot)$) SoundAQnet, and the language model ($lm(\cdot)$). 
SoundSCaper aims to generate a language description $\boldsymbol{D}$ to describe the soundscape based on the input audio clip $\boldsymbol{A}$. 
To this end, first, we extract the AS and AE information, as well as the affective response values of eight AQs in Fig. \ref{SCM}, i.e. \textit{pleasant}, \textit{eventful}, \textit{chaotic}, \textit{vibrant}, \textit{uneventful}, \textit{calm}, \textit{annoying}, and \textit{monotonous}
\cite{araus} from the audio clip $\boldsymbol{A}$, 
by building an acoustic model ($am(\cdot)$), i.e., $\{AS, AE, AQ\} = am(\boldsymbol{A})$. Then, we form a textual description of the soundscape by a language model like GPT \cite{openai_chatgpt} ($lm(\cdot)$), i.e., $\boldsymbol{D} = lm(AS, AE, AQ)$. 


\vspace{-1mm}
\section{The proposed SoundSCaper system}\label{LLM_SSD_framework}
\vspace{-1mm}
This section introduces SoundSCaper in detail from two aspects: the acoustic model and the language model.

\vspace{-3mm}
\subsection{Acoustic Model: the Proposed SoundAQnet}\label{acoustic_model} 
 
The SoundAQnet aims to simultaneously model ASs and AEs in acoustic environments, as well as the corresponding affective responses to the soundscape, i.e., PAQ 8 attributes shown in Fig. \ref{SCM}. 
\mybluehl{In recognition of AS and AE, the  Mel spectrogram is a typical acoustic feature \cite{pann}\cite{taslp_cSEM} with excellent performance.}
In soundscape studies \cite{erfanian2021psychological}\cite{erfanian2019psychophysiological}\cite{AI_soundscape}, 
\mybluehl{perceived loudness, approximated as calculated as Zwicker-loudness in ISO 532-1:2017 \cite{ISO_loudness}, is of primary importance to estimate AQ even if AEs are known.}
Thus, both Mel and loudness are used to capture the AS, AE, and AQ in soundscape audio clips.
\mybluehl{The novelty of SoundAQnet lies in the design of the network structure and the loss function to enable the joint modelling of ASs, AEs, and AQs, as described below.}

\subsubsection{\mybluehl{\textbf{Network}}}


\mybluehl{To process these long input features with few parameters, SoundAQnet uses dilated convolution \cite{HDC_dilated} in its Mel-based and Loudness-based branches to obtain a larger receptive field size (RFS) with limited computing resources. }


\begin{figure*}[b] 
	\setlength{\abovecaptionskip}{-2mm}  
    \setlength{\belowcaptionskip}{-2cm}   
    \vspace{-4mm}
	\centerline{\includegraphics[width = 0.85 \textwidth]{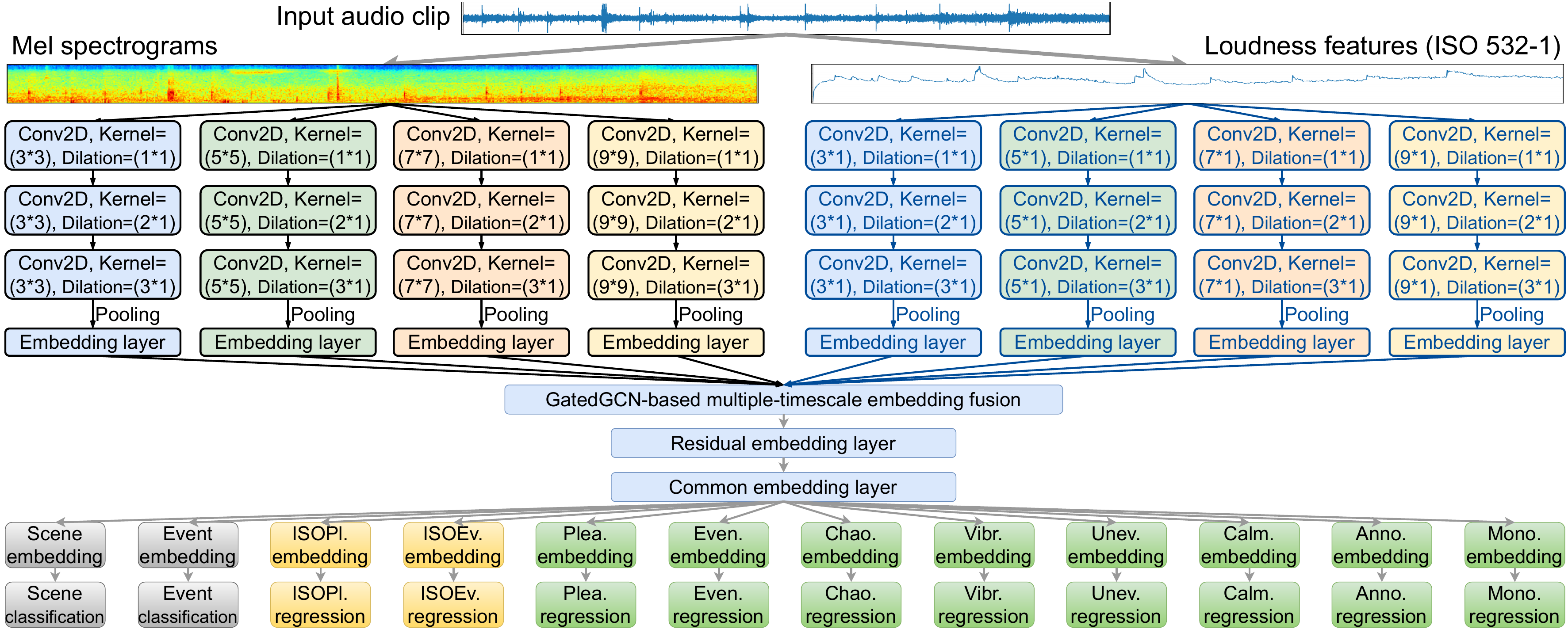}}
	\caption{The SoundAQnet simultaneously models acoustic scene (AS), audio event (AE), and emotion-related affective quality (AQ).}
	\label{SoundAQnet}
\end{figure*}

\textit{1.1) \textbf{Mel-based Branch}:} 
\mybluehl{Audio events are often in different spectral-temporal scales, ranging from short and transient to long-lasting events.} 
Therefore, the Mel-based branch employs four sub-branches to obtain multi-scale representation, by applying convolutional kernels of different sizes, i.e., [(3, 3), (5, 5), (7, 7), (9, 9)], to the input features on the (time, frequency) axis, respectively. Each sub-branch comprises three convolution blocks with 16, 32, and 64 filters.  
\mybluehl{
The gridding artifacts \cite{gridding} of dilated convolutions can lead to compromised information continuity and loss of local information.
Thus, a hybrid dilated convolution \cite{HDC_dilated} scheme is used, where the dilation rates in the three convolutional blocks are in order [(1,1), (2,1), (3,1)], allowing the branches to extract context from a broader and more coherent receptive field along the time axis.} 
The dilation rate only varies along the time axis, as the frequency dimension is often relatively small.

In the Mel branch, each 2D convolution (Conv2D) block refers to the design of VGG \cite{vgg} and consists of two convolution layers.
Taking the largest kernel (9, 9) as an example, there are three Conv2D blocks, i.e., six 2D convolution layers, according to the convolution RFS calculation formula, 
\begin{equation}\label{formula_receptive_field}
\setlength{\abovedisplayskip}{1pt}
\setlength{\belowdisplayskip}{1pt}
F_i=(F_{i-1}-1) \times stride + k
\end{equation} 
where $F_i$ denotes the $i$-th convolution layer's RFS relative to the input feature map, $F_0=1$, $k$ is the convolution kernel size, and $stride$ defaults to 1. 
If there is no pooling operation, according to Eq.~(\ref{formula_receptive_field}), in the first Conv2D block, the RFS of the first layer on the time axis is $F_1=9$, and that of the second layer is $F_2=17$. For the dilated convolution, the RFS is
\begin{equation}
\setlength{\abovedisplayskip}{1pt}
\setlength{\belowdisplayskip}{1pt}
F_i=(F_{i-1}-1) \times stride + k + (k-1)(r-1)
\end{equation}
where $r$ is the dilation rate.
\mybluehl{For the second Conv2D block with dilation rate (2, 1), on the time axis, $F_3=33$, and $F_4=49$. For the third Conv2D block with dilation rate (3, 1), on the time axis, $F_5=73$, and $F_6=98$.
With these Conv2D blocks without pooling, SoundAQnet requires the input features to be at least 98 frames long. With the frame hop of 10\textit{ms}, the corresponding input clip length is at least 980\textit{ms}.}
It is challenging to identify AS or AE from 1-second audio clips, even for humans, let alone the 8D AQs. Furthermore, if pooling is not used in Conv2D, it will increase the parameters and the computation load. After comprehensive trade-offs, we add pooling operations to these multiscale Conv2D blocks, resulting in a minimum input audio length of 2.80\textit{s}.

After the last Conv2D block of each sub-branch, global pooling unifies the length of multiscale representations. These dimension-unified representations are fed into separate embedding layers to output 64-dimensional embeddings for fusion.


 
\textit{1.2) \textbf{Loudness-based Branch}:} 
The loudness of AEs affects how humans perceive them, thereby affecting human-perceived AQ. Many models for noise annoyance within one class of sounds (e.g., road traffic) even rely exclusively on loudness. 
\mybluehl{The Mel spectrogram can estimate the loudness changes to some extent.
However, since perceived loudness has been studied in psychoacoustics for decades, it is more effective to directly use loudness as a psychoacoustic feature in the model instead of estimating it implicitly with the model.
Loudness is introduced as a 1D feature with the unit of \textbf{sone}.
In SoundAQnet, loudness features are extracted after calibration with a reference signal, i.e. a sine wave of 1kHz at 60dB, from the ISO 532-1 standard \cite{ISO_loudness}. 
Given $N$ frames, the size of loudness features is ($N$, 1). Since the generation, development, and fading of emotions is a dynamic process with different time scales of importance, we also use multiscale convolution blocks in the loudness branch to extract the PAQs defined in ISO/TS 12913-3:2019 \cite{ISO_soundscape_PAQ} shown in Fig. \ref{SCM}.} More specifically, the multiscale convolution kernels used in the sub-branch are of dimension [($3$, 1), ($5$, 1), ($7$, 1), ($9$, 1)].
The remaining part of the loudness branch is the same as that in the Mel branch.


\textit{1.3) \textbf{Graph-based Multiscale Embedding Fusion}:} 
To fuse representations from Mel- and Loudness-based branches, we consider the representation embeddings as node features and build a fully connected soundscape-dependent multiscale sound-AQ representation graph. Here, our hypothesis is that since the model is trained with co-supervised labels of AS, AE, and AQ in the soundscape, the sound-AQ representation graph will automatically couple the acoustic environment and AQ while updating node features and learning relationships between nodes with different time granularities. 
\mybluehl{
That is, by updating the features of edges connecting nodes, the message about the differences between different timescale nodes is passed to each other through edges in the graph, thereby further aggregating and fusing information from different scales. Thus, it is crucial to learn edge features in the sound-AQ representation graph during updating.  
Then, we use the gated graph convolutional network (GatedGCN) \cite{gatedGCN} in the graph-based multiscale embedding fusion layer, where the node and edge features are updated simultaneously.}
Once the soundscape-dependent sound-AQ representation graphs containing $n$ nodes and $n\times n$ edges are obtained, we use one layer of GatedGCN to model these graphs. The number of nodes is $n=8$; the size of each node embedding is 64.

\textit{1.4) \textbf{Co-embedding and Separate Embedding Layers}:} 
To absorb the information updated by the sound-AQ graph while considering its original acoustic representations, we residually connect \cite{he2016deep} the node embeddings output by the graph with the input to the graph. We concatenate all embeddings and input them into the common embedding layer to learn the acoustic context- and AQ-related embeddings. This allows the classification and regression tasks to use all the information within the common embedding captured by the model. Next, separate embedding layers are used for the ASC and AEC tasks, and the human-perceived AQ regression tasks, to learn representations for each target individually.

\subsubsection{\textbf{Loss Functions}} 
The SoundAQnet involves 2 classification objectives (AS, AE) and 10 regression objectives (\textit{ISOP}, \textit{ISOE}, 8D AQs).
ISO Pleasantness (\textit{ISOP}) and ISO Eventfulness (\textit{ISOE}) in Fig. \ref{SCM}, 
can be calculated as follows:

\vspace{-2mm}
\begin{equation}
\setlength{\abovedisplayskip}{0pt}
\setlength{\belowdisplayskip}{0pt}
ISOP = k^{-1} (\sqrt{2} r_{pl} - \sqrt{2} r_{an} + r_{ca} - r_{ch} + r_{vi} - r_{mo})
\vspace{-0mm}
\end{equation} 
\begin{equation}
ISOE = k^{-1} (\sqrt{2} r_{ev} - \sqrt{2} r_{ue} - r_{ca} + r_{ch} + r_{vi} - r_{mo})
\vspace{-1mm}
\end{equation}  
where $r_{\{pl, ev, ch, vi, ue, ca, an, mo\}}$ $\in$ $\{1,2,3,4,5\}$ are human response values to 8D AQs: \textit{pleasant}, \textit{eventful}, \textit{chaotic}, \textit{vibrant}, \textit{uneventful}, \textit{calm}, \textit{annoying}, and \textit{monotonous}, respectively, and $k=8+\sqrt{32}$. \textit{ISOP} and \textit{ISOE} are related to AQs, so the model's prediction for \textit{ISOP} can imply the overall performance of human-perceived AQ predictions.

\mybluehl{
\textit{2.1) \textbf{AS}:}  
For the ASC tasks, cross entropy (CE) \cite{taslp_cSEM} is used as the loss function that measures the difference between the prediction $p_{s}$ and its label ${y}_{s}$, i.e. $\mathcal{L}_\text{1} = CE(p_{s}, y_{s})$.
}

\mybluehl{
\textit{2.2) \textbf{AE}:}  
For the AEC tasks, binary cross entropy (BCE) is used as the loss function that measures the difference between the prediction $p_{e}$ and its label ${y}_{e}$, i.e. $\mathcal{L}_\text{2} = BCE(p_{e}, y_{e})$.
}

\mybluehl{
\textit{2.3) \textbf{AQ}:}  
For the AQ regression tasks, mean squared error (MSE) is used as the loss function. Specifically, $\mathcal{L}_\text{3} = MSE(p_{isop}, ISOP)$ and $\mathcal{L}_\text{4} = MSE(p_{isoe}, ISOE)$, where $p_{isop}$ and $p_{isoe}$ are predictions of \textit{ISOP} and \textit{ISOE}, respectively. 
Also, $\mathcal{L}_\text{n} = MSE(p_{aq}, y_{aq}), n \in [5, 12]$,
where $p_{aq}$ and ${y}_{aq}$ are predictions and labels of each type of AQ in 8D AQs. 
}

\mybluehl{
\textit{2.4) \textbf{Total Loss}:}  
Optimizing the 12 objectives with 12 losses is challenging. 
Typical Pareto optimization \cite{lin2019pareto} is unsuitable for SoundAQnet. Because the quantification of AQ has a certain degree of ambiguity, assuming that 3$\pm$0.25$\approx$3 for $r_{pl}$, its prediction $\pm$0.1 has little impact on the final AQ output.
Hence, compared to emotion-related AQs, SoundAQnet needs to perform better in ASC and AEC with explicit classification goals, i.e., SoundAQnet does not aim to achieve the Pareto optimality of all 12 objectives. 
Human perception times for various scenes, events, and emotions may vary.  
This implies that different learning rates might be better suited for optimizing the 12 classification and regression losses.}
Hence, GradNorm-like optimizations \cite{chen2018gradnorm}, which aim to learn multiple tasks at a similar rate from a gradient view, do not suit SoundAQnet. 
After considering the computational effort and training speed, we choose uncertainty-based weighting \cite{kendall2018multi} to fuse the 12 losses.  
\begin{equation}
\mathcal{L}= \sum\nolimits_{i=1}^{2}(\frac{1}{{\sigma_i}^2}{\mathcal{L}}_{i} + \text{log} {\sigma_i}) + \sum\nolimits_{j=3}^{12}(\frac{1}{{2\sigma_j}^2}{\mathcal{L}}_{j} + \text{log} {\sigma_j})
\vspace{-1mm}
\end{equation}  
where the learnable noise parameter $\sigma$ denotes the uncertainty associated with the corresponding task \cite{kendall2018multi}, and the logarithmic function based penalty term prevents $\sigma$ from becoming excessively high. The higher the uncertainty $\sigma$, the lower the contribution of the loss to the total loss.

\vspace{-2mm}
\subsection{Language Model: Customised LLM for Caption Generation}
\label{sec::LLM} 
This section explains how the AS, AE, and AQ information, obtained by SoundAQnet, can be turned into captions to describe the soundscape in human-understandable language by customizing an LLM with prompt engineering techniques.

\begin{figure}[b] 
	\setlength{\abovecaptionskip}{-1mm}  
    \setlength{\belowcaptionskip}{0cm}   
    \vspace{-4mm}
    \centerline{\includegraphics[width = 0.5 \textwidth]{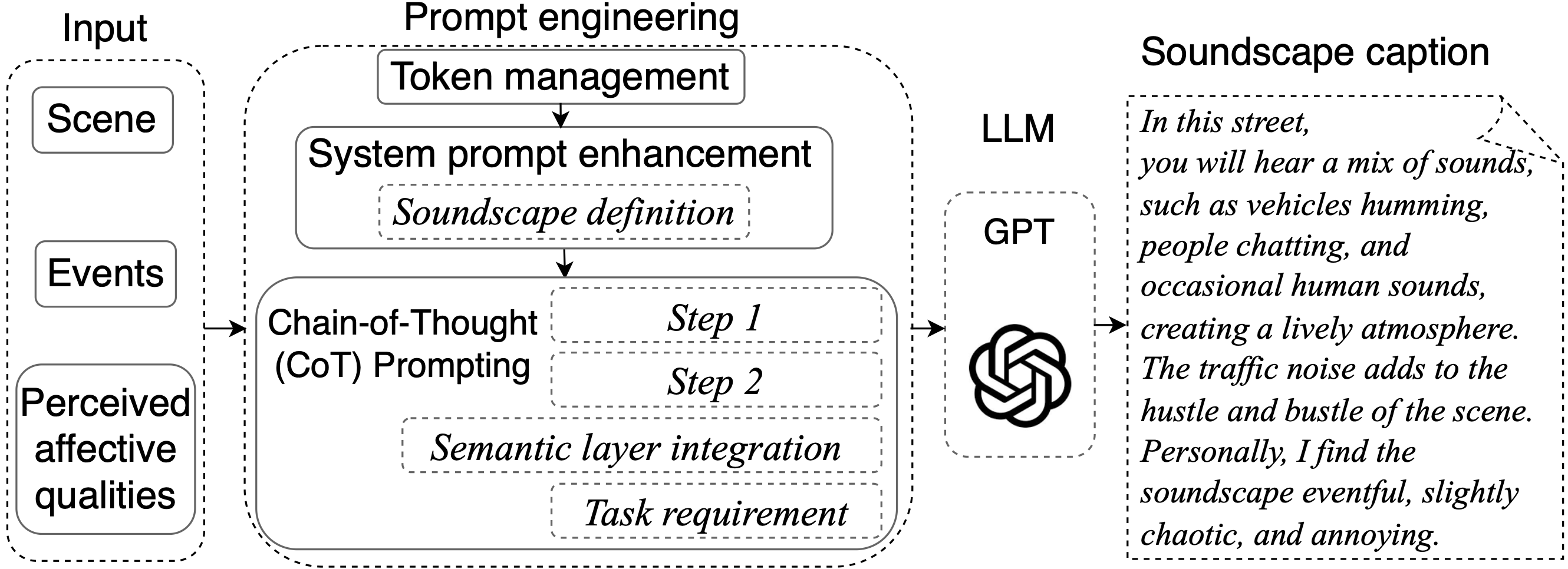}}
	\caption{Process of the LLM part in the proposed SoundSCaper system.}
	\label{LLM_GPT}
\end{figure}

\subsubsection{\textbf{Pipeline for Caption Generation}}

 As shown in Fig. \ref{LLM_GPT}, we design a pipeline to convert the numerical information of AS and AE and the emotion-relevant AQ into a textual description of the soundscape, with the help of the prior knowledge learned by an LLM. There are various methods for integrating acoustic information into LLM, including early, mid, and late fusion \cite{chen2024its}, corresponding to information fusion at the feature, intermediate representation, and decision level, respectively. In this work, we choose GPT as the LLM to fuse the decision results of scenes, events, and affective qualities parsed by SoundAQnet, i.e. $\boldsymbol{D} = lm(AS, AE, AQ)$ as discussed in Section \ref{Soundscape_caption}, where $lm$ represents the GPT model, such as DaVinci, GPT-3.5-Turbo, and GPT-4, according to OpenAI \cite{openai_chatgpt} services. We choose GPT-3.5-Turbo, which offers a tradeoff among generation accuracy, response speed, number of tokens, and cost. An advantage of this pipeline is that SoundSCaper's pipeline can directly use general LLMs such as GPT and potentially benefit from their future updates.

In the SoundSCaper system, LLM serves two purposes: (1) to present knowledge extracted from audio recordings using SoundAQnet in fluent language; (2) to introduce common knowledge on expected AEs and AQs in a specific context, that is, a specific AS. \mybluehl{Currently, there is no large-scale dataset in the soundscape domain that pairs audio with affective descriptive captions encompassing AE, AS, and human-perceived AQ. As a result, existing resources are inadequate for training audio-to-text soundscape-language models or for fine-tuning LLMs. In addition, fine-tuning LLMs on expert-annotated affective captions may introduce biases related to individual personalities and cultural perspectives. Moreover, fine-tuning LLM may increase the chance of hallucination \cite{gekhman2024does}.}
For these reasons, 
SoundSCaper uses a general pre-trained GPT instead of fine-tuning a model.
This choice has several advantages: 
\mybluehl{
it avoids the dilemma of no paired data of sound and affective description to support the training or fine-tuning of soundscape LLM, and also avoids time-consuming and laborious training;} directly using generic LLMs allows the acoustic and language models to be updated separately; the system can be extended with non-acoustic contextual information, if available; and captions can be easily provided in other languages.

\subsubsection{\textbf{Customized LLM}}

\mybluehl{As a novel design for soundscape caption generation, we integrate prompt engineering to enhance the output's contextual accuracy, affective depth, and narrative clarity to decrease hallucinations, as shown in Fig. \ref{LLM_GPT}. 
The employed prompt engineering strategies include structured output generation, system prompt enhancement, semantic layer integration, and chain-of-thought (CoT) prompting, which are more effective than fine-tuning LLM in improving model performance and reducing hallucinations \cite{rumiantsau2024beyond}.}

\textit{2.1) \textbf{Token Management}:}  
To optimize the amount of input and output of the LLM, tokens are managed. For an audio clip, the scene, AS, is unique, but multiple AEs may be detected simultaneously. 
Hence, to process AE probabilities predicted by SoundAQnet, we first use an empirical probability threshold of 0.3 to obtain the text labels of AEs present in the audio clip. 
Then, for AQs predicted by SoundAQnet, we prioritize input tokens with strong responses, that is, high predicted values. This is because human attention is often attracted by the dominant AE while being influenced by the dominant AQ. \mybluehl{The ASSC task aims to describe the most relevant information in soundscapes. Additionally, we instruct the LLM to limit the generated descriptions to 200 tokens.} These strategies manage the consumption of input and output tokens.

\textit{2.2) \textbf{System Prompt Enhancement}:} 
We use the soundscape definition from ISO 12913-1:2014 \cite{ISO_soundscape} to provide LLM with a conceptual framework that incorporates perception (psychology) and understanding (cognition) in the description, as well as context, people, and society. \mybluehl{
In addition, the prompt asks the LLM to define its role as a soundscape expert to fit the soundscape research field.}

\mybluehl{
\textit{2.3) \textbf{Semantic Layer Integration}:} 
Rule-based semantic constraints are used to reduce hallucinations \cite{rumiantsau2024beyond} in ASSC. The LLM is instructed to use only the provided AE labels and avoid interpretive or embellished descriptions. For example, \textit{vehicle} should not be expressed as `the hum of engines and honking of horns'; LLM should directly list the AEs. These constraints help ground the output, improving accuracy and consistency without requiring LLM fine-tuning.
}

\textit{2.4) \textbf{Chain-of-thought Prompting}:} 
\mybluehl{This part guides the LLM through a logical analysis sequence for structured output generation, from AS and AE classification to AQ regression. 
The structured approach aids in systematically tackling complex auditory and affective analyses.}
The task is decomposed into focused subtasks to help LLM understand the relationship between input acoustic environment information and AQs based on its large-scale prior knowledge to ensure comprehensive and accurate caption generation. Prompts are as follows:

\begin{fontsize}{8pt}{10pt}    
\textit{\textrm{...As an expert in soundscape analysis, your task is ...}}
\end{fontsize}

\begin{fontsize}{8pt}{10pt}  
\textit{\textrm{Step 1: According to the events and their corresponding probability ... in this scene, identify ... and describe the auditory scenario...}}
\end{fontsize}

\begin{fontsize}{8pt}{10pt}   
\textit{\textrm{Step 2: Describe your feelings based on the ratings on this soundscape......}}
\end{fontsize}


Full prompts and scripts can be found in the \textbf{\textit{homepage}}.

\section{\mybluehl{Acoustic model (SoundAQnet)} experiment}\label{AM_section_setup}


\subsection{Dataset}\label{acoustic_dataset}

Commonly used large-scale audio datasets like AudioSet \cite{audioset} and FSD50K \cite{fonseca2021fsd50k} do not contain corresponding ``\textit{subjective}" labels regarding the PAQ of recording environments \cite{araus}, which prevents them from being used to train SoundAQnet. To the best of our knowledge, the recently published ARAUS dataset \cite{araus} is the largest soundscape dataset with the most complete human affective responses to AQs. Therefore, the ARAUS dataset is used to train the proposed SoundAQnet.

\mybluehl{ARAUS contains 25440 30-second binaural audio clips, totaling 212 hours. With the efforts of 605 participants}, each audio clip has 8D AQ values annotated according to ISO/TS 12913-2 \cite{ISO_soundscape_data}. ARAUS is augmented on the Urban Soundscapes of the World (USotW) \cite{usotw} dataset. Each augmented soundscape is made by digitally adding maskers (\textit{birds}, \textit{water}, \textit{wind}, \textit{traffic}, \textit{construction}, or \textit{silence}) to an urban soundscape recording at soundscape-to-masker ratios \cite{araus}. The maskers are AEs. Hence, ARAUS meets the needs of SoundAQnet training with affective supervision information. Unfortunately, ARAUS does not have AS and AE labels.

The USotW \cite{usotw}, as synthesis material for ARAUS, contains 360-degree video clips with GPS locations, which allows us to easily identify their corresponding scene. There are 3 classes of ASs, namely \{\textit{public square}, \textit{park}, \textit{street traffic}\}. Following ARAUS synthesis rules, we obtain the AS labels of ARAUS.

Although six types of AEs have been explicitly added in ARAUS, we cannot directly use the six labels as AE labels because USotW already contains numerous AEs. 
To obtain the detailed AE labels in ARAUS, we first use the excellent pre-trained audio model PANNs \cite{pann} to label each audio clip with a one-second-level pseudo-label. Since the PANNs model is trained on AudioSet, a large-scale dataset with 527 classes of AEs, each one-second audio clip is assigned with a 527-dimensional soft pseudo label, corresponding to the probability of 527 classes of AEs within this second.  
Then, the soft pseudo-labels are mapped into hard pseudo-labels consisting of \{0, 1\} by a threshold of 0.5.
After accumulating and sorting the hard pseudo-labels for all one-second segments, we obtain the number of occurrences for the 527 classes of AEs in ARAUS, ranked from high to low. After considering the six types of AEs added in ARAUS, a total of 15 AE labels are obtained, which are \mybluehl{\{\textit{Bird}, \textit{Animal}, \textit{Wind}, \textit{Water}, \textit{Natural sounds}, \textit{Vehicle}, \textit{Traffic}, \textit{Sounds of things}, \textit{Environment and background}, \textit{Outside, rural or natural}, \textit{Speech},  \textit{Human sounds},  \textit{Music}, \textit{Noise}, \textit{Silence}\}}.  
For training SoundAQnet, only clip-level AE labels are needed to distinguish whether the target AE is within the input clip. Hence, we again use PANNs to label the clip-level 527 AE probabilities for each audio clip. Then, the probabilities of 15 classes of target AEs are taken out and binarized into hard labels using a threshold of 0.1.

\mybluehl{In the experiment of ARAUS \cite{araus}, the validation set has 5040 samples, while the test set has only 48 samples.} 
The test set may be too small to effectively evaluate the performance of our model. Thus, we randomly shuffled and re-divided the ARAUS dataset. In proportion, 19152 30-second audio clips are randomly selected as the training set; 2520 and 3576 audio clips are chosen as the validation and test sets, respectively. \mybluehl{To avoid intersections between the three sets, the number of audio clips used in this paper is 25248, rather than 25440.}

\vspace{-3mm}
\subsection{Experimental Setup of Acoustic Model}

\textbf{Mel feature.} 
The setting of log Mel features follows that of PANNs \cite{pann}. The 64 Mel bins are extracted by STFT with a Hamming window length of 32\textit{ms} and a hop size of 10\textit{ms}, resulting in Mel features having 3000 frames. 

\textbf{Loudness feature.}
Loudness features are extracted directly using the \textit{ISO\_532-1.exe} loudness program\protect\footnote{https://standards.iso.org/iso/532/-1/ed-1/en} recommended by ISO 532-1:2017 standard (Zwicker method) \cite{ISO_loudness}. The input audio files are calibrated with a \textit{``.wav''} file containing the calibration signal, which is a sine wave at 1 kHz 60 dB. Then, the loudness features are calculated in frames of 2\textit{ms}, resulting in Zwicker-loudness with 15000 frames.
We upload the modified Python code and files to the \textbf{\textit{homepage}}.

\textbf{Training settings.} 
Adam optimizer is used to minimize the loss, with a learning rate 5e-4 and batch size 32.  
Since SoundAQnet contains a total of 12 tasks for classification and regression, referring to the settings in ARAUS \cite{araus}, this paper monitors the \textit{ISOP} loss on the validation set in early stopping. Starting from the 10th epoch, if the validation loss value of \textit{ISOP} does not decrease within 10 epochs, training is stopped. The model is trained for a maximum of 100 epochs. The model is trained 10 times without a fixed seed to obtain the mean performance over the 10 runs. 
Accuracy (Acc) and threshold-free AUC \cite{taslp_cSEM} are used to evaluate ASC and AEC results. The mean squared error (MSE) is used to measure the regression results. 
The AS and AE labels that we annotated for ARAUS, code, and trained models are all available on the \textbf{\textit{homepage}}.




\vspace{-2mm}
\subsection{\mybluehl{Comparisons to Other Models}}\label{RQ4}

\vspace{-4mm}
\begin{table}[H] 
	\setlength{\abovecaptionskip}{-1mm}   
	\setlength{\belowcaptionskip}{-0.2cm}  
	\renewcommand\tabcolsep{1pt} 
	\centering
	\caption{\mybluehl{Comparison of different models on the test set (batch size=32)}.}
	\begin{tabular}{  
	p{0.15cm}<{\centering}|
	p{2.1cm}<{\centering}|
    p{0.7cm}<{\centering}|
    p{1.0cm}<{\centering}|
    p{1.1cm}<{\centering}|
	p{0.9cm}<{\centering}|
	p{0.5cm}<{\centering}|
 p{1.65cm}<{\centering}
 }
	
		\toprule[1pt] 
		\specialrule{0em}{0.1pt}{0.1pt}
 
\multirow{2}{*}{\#} &
\multirow{2}{*}{Model} &  Param & \multicolumn{2}{c|}{Inference} &
\multicolumn{1}{c|}{ASC} & \multicolumn{1}{c|}{AEC} &{AQ regression}\\
		\cline{4-8}     
	    &  & (M) & time(ms) & GPU(GB) & \textit{Acc.}(\%) & \textit{AUC} & \text{MSE Mean} \\
	\hline 

    1 &  AD\_CNN \cite{araus} & 0.52 & 5.6 & 1.79 & 89.63  &  0.84   &  1.128 \\   
    2 &  Baseline\_CNN & 1.01 & 4.3 & 1.18 & 87.87  &  0.92   &  1.315  \\   
    3 &  Hierarchical\_CNN & 1.01 & 4.6 & 1.18 & 89.82  &  0.89    &  1.293  \\
 
    4 &  MobileNetV2 & 2.26 & 5.5 & 1.94 & 89.67    &  0.92  & 1.145 \\   

    5 &  YAMNet  & 3.21 & 4.9 & 2.16 & 88.84 &  0.90    &  1.199 \\   

    6 &  CNN-Transformer & 12.29 & 4.4 & 1.24 & 92.80  &  0.93    &  1.339  \\  
    
    7 &  PANNs \cite{pann} & 79.73 & 19.0 & 5.67 & 93.57  &  0.90  & 1.156 \\    
    \hline 
    
    8 &  SoundAQnet & 2.70 & 18.9 & 3.22 & \textbf{95.31}   &  \textbf{0.94}  & \textbf{1.054}   \\  
    
		\specialrule{0em}{0pt}{0em}
		\bottomrule[1pt]
	\end{tabular}
	\label{tab:other_models}
\end{table}

 

\vspace{-1mm}
There are no other models similar to SoundAQnet for simultaneously modelling the AS, AE, and emotion-related AQ. Previous studies on AQ in soundscapes often use traditional linear regression to predict some AQ response values, while recent deep-learning-based studies only focus on a few specific AQs \cite{hou23_interspeech}\cite{lunden2016urban}\cite{soundscape_IADS}.
Therefore, we compare SoundAQnet with deep-learning models that perform well for auditory scene and event analysis, i.e., ASC and AEC tasks, as shown in Table \ref{tab:other_models}.

In Table \ref{tab:other_models}, \#1 refers to the CNN used in the ARAUS paper \cite{araus}.  
CNN in \#2 is the baseline for benchmarking the multiscale convolution-based SoundAQnet. It consists of 4 convolutional layers, each with 16, 32, 64, and 128 filters, and their corresponding kernel sizes of 3, 5, 7, and 9, respectively. After the convolutional layers, there are parallel ASC and AEC layers and regression layers for AQs. 
Hierarchical CNN in \#3 aims to identify AS based on the predictions of AE, exploiting the implicit hierarchical relationship between AS and AEs \cite{taslp_cSEM}. 
Compared with \#2, the ASC in \#3 is better, but the AEC is affected by the 
hierarchical relationship.
MobileNetV2 in \#4 is a lightweight CNN that uses depthwise separable convolution to reduce the computational cost \cite{sandler2018mobilenetv2}. YAMNet in \#5 is a CNN-based baseline for AEC from Google. Given the excellent performance of Transformers on audio tasks \cite{taslp_cSEM}, \#6 proposes CNN-Transformer, an encoder from Transformer \cite{Transformer} is added after the convolutional layer in Baseline CNN, to combine the spatial feature extraction capability of CNN with the temporal modelling capability of Transformer. 
Compared with \#2, the introduction of Transformer encoder in \#6 enhances the model's ability to discriminate acoustic scenes and events, and improves its classification results, but its overall result on 8D AQ regressions is not as good as those of the pure CNN in \#4. The reason may be that, compared with Transformer encoder modelling AQs from the hidden layer features with the global perspective, CNN relies on a fixed-size convolutional kernel and performs better in learning the hidden layer features from different local perspectives, which is beneficial for modelling unique representations of each AQ.

Overall, the proposed SoundAQnet, which simultaneously models AS, AE, and human-perceived AQ, achieves the best results in ASC, AEC, and affect-related regression tasks with a similar number of parameters as MobileNetV2. 
\mybluehl{More information on computational efficiency, such as training/inference speed, GPU, and hardware requirements, see the \textbf{\textit{homepage}}.}

\vspace{-2mm}
\subsection{\mybluehl{Ablation Studies of SoundAQnet}}\label{RQ1_3} 


\noindent
\textit{\mybluehl{1) Ablation study on acoustic features}}

Tables \ref{tab:ablation_two_features_ASC_AEC_ISO} and \ref{tab:ablation_two_features_PAQ8_part1} present the performance of SoundAQnet on ASC, AEC, ISO Pleasantness (\textit{ISOP}) and ISO Eventfulness (\textit{ISOE}) regression, and emotion-related AQ regression tasks when using different acoustic features, respectively. 
They are the mean results over 10 runs. 
When using single-class acoustic features, SoundAQnet retains only the corresponding branches, and the number of nodes in the graph-based fusion layer is reduced by half, i.e., $n=4$.

\vspace{-4mm}
\begin{table}[H] 
	\setlength{\abovecaptionskip}{-1mm}   
	\setlength{\belowcaptionskip}{-0.2cm}  
	\renewcommand\tabcolsep{1pt} 
	\centering
	\caption{Mean performance of SoundAQnet on the test dataset (part 1).}
	\begin{tabular}{  
	p{0.3cm}<{\centering}|
	p{0.7cm}<{\centering}
 p{1.3cm}<{\centering}|
	p{1.1cm}<{\centering}|
	p{0.9cm}<{\centering}|
 p{0.9cm}<{\centering}|
 p{0.9cm}<{\centering}|
 p{1.cm}<{\centering}|
 p{1.cm}<{\centering}}
	
		\toprule[1pt] 
		\specialrule{0em}{0.1pt}{0.1pt}
 
\multirow{2}{*}{\#} &
\multicolumn{2}{c|}{Acoustic feature} & \multicolumn{1}{c|}{ASC} & \multicolumn{1}{c|}{AEC}  & \multicolumn{1}{c|}{\textit{ISOP}} & \multicolumn{1}{c|}{\textit{ISOE}} & \textit{pleasant} & \textit{eventful} 
 \\
		\cline{2-9}     
	       & Mel & Loudness & 
\textit{Acc.} (\%) & \textit{AUC}
 & \multicolumn{4}{c}{\textit{MSE}}\\
	\hline 

    1 &  \XSolidBrush & \CheckmarkBold & 73.61  &  0.868   &  0.116  & 0.129 & 0.993 & 1.161 \\  
    2 &  \CheckmarkBold & \XSolidBrush  & 94.07  &  0.934   &  0.112  & 0.116 &  0.943 & 1.093 \\   
        
    3 &  \CheckmarkBold & \CheckmarkBold  & \textbf{95.31}  &   \textbf{0.941}   &  \textbf{0.106}   &  \text{0.115} &  \textbf{0.899}   &  \text{1.068}   
\\  
  
		\specialrule{0em}{0pt}{0em}
		\bottomrule[1pt]
	\end{tabular}
	\label{tab:ablation_two_features_ASC_AEC_ISO}
\end{table}

\vspace{-8mm}
\begin{table}[H] 
	\setlength{\abovecaptionskip}{-1mm}   
	\setlength{\belowcaptionskip}{-0.2cm}  
	\renewcommand\tabcolsep{1pt} 
	\centering
	\caption{Mean performance of SoundAQnet on the test dataset (part 2).}
	\begin{tabular}{  
	p{0.3cm}<{\centering}|
	p{0.5cm}<{\centering}
 p{0.8cm}<{\centering}|
	p{0.9cm}<{\centering}|
	p{0.9cm}<{\centering}|
 p{1.3cm}<{\centering}|
 p{0.8cm}<{\centering}|
 p{1.2cm}<{\centering}|
 p{1.4cm}<{\centering}}
	
		\toprule[1pt] 
		\specialrule{0em}{0.1pt}{0.1pt}
 
\multirow{2}{*}{\#} &
\multirow{2}{*}{\text{Mel}} & 
\multirow{1}{*}{\text{Loud-}} & \textit{chaotic} & \textit{vibrant}  & \textit{uneventful} & \textit{calm} & \textit{annoying} & \textit{monotonous}
\\
		\cline{4-9}     
         &  & {\text{ness}} & \multicolumn{6}{c}{\textit{MSE}}\\
	\hline  
    1 &  \XSolidBrush & \CheckmarkBold &  1.187 & 1.067   &  1.237  &  	1.105	  &  1.191  &  	1.234
 \\  
    2 &  \CheckmarkBold & \XSolidBrush  &  1.098  &  	0.975  &  	1.165  &  	1.043  &  	1.105  &  	1.167  \\ 
    3 & \CheckmarkBold & \CheckmarkBold  & \textbf{1.079}  &   \text{0.979}   & \text{1.168}  &  \textbf{0.999}  & \textbf{1.083}  &  \text{1.159}   
    \\  
		\specialrule{0em}{0pt}{0em}
		\bottomrule[1pt]
	\end{tabular}
	\label{tab:ablation_two_features_PAQ8_part1}
\end{table}

\vspace{-2mm}
The comparison of \#1 and \#2 in Table \ref{tab:ablation_two_features_ASC_AEC_ISO} shows that for the acoustic environment-related ASC and AEC, the Mel feature is more effective than the loudness feature, consistent with previous research \cite{AI_soundscape}. The reason for this is easy to understand, following the notations in Section \ref{acoustic_model}, Mel features with dimension ($N$, 64) include spectral information, which is essential for recognizing sounds, while this essential information is lost in over-compressed loudness features with dimension ($N$, 1).  
Compared with Mel-based \#2, the fused feature in \#3 performs better in the regression of \textit{ISOP}, \textit{pleasant}, \textit{chaotic}, \textit{calm}, and \textit{annoying}, as well as in classifications of AS and AE. 
\mybluehl{This indicates that introducing Zwicker-loudness 
explicitly can help Mel-based SoundAQnet on partial AQ regressions, and AS and AE classification. The effect of adding loudness is stronger for \emph{ISOP} (the valence axis in Fig. \ref{SCM}), which is in line with expectations \cite{erfanian2019psychophysiological}\cite{ISO_loudness}.}
Since \textit{ISOP} and \textit{ISOE} are linear combinations of the 8D AQs and thus do not offer additional insight, we will omit their results in the following experiments due to space limitations.



\noindent
\textit{\mybluehl{2) Ablation study on multiscale sub-branches}}

The duration of AEs may vary between a few tens of milliseconds, such as bird chirps, and several minutes, such as music.  
\mybluehl{AQs along the valence dimension, \emph{ISOP}, may be determined by short sounds that are identified as calming or annoying. AQs that are aligned with high arousal, high \emph{ISOE}, such as \emph{eventful}, \emph{chaotic}, and \emph{vibrant}, are related to multiple changes in the sound environment and require longer time windows to be identified. This ablation study aims to explore whether these expected relationships can be linked to the kernel sizes used in different branches.}

\begin{table*}[t] 
	\setlength{\abovecaptionskip}{-1mm}   
	\setlength{\belowcaptionskip}{0cm}  
	\renewcommand\tabcolsep{1pt} 
	\centering
 \vspace{0mm}
	\caption{Mean performance of 10 runs of SoundAQnet with convolution branches of different kernel sizes on the test set.}
	\begin{tabular}{  
	p{0.3cm}<{\centering}|
 p{0.4cm}<{\centering}
 p{0.4cm}<{\centering}
 p{0.4cm}<{\centering}
 p{0.4cm}<{\centering}|
    p{2cm}<{\centering}| 
    p{0.7cm}<{\centering}|
    p{1.1cm}<{\centering}|
	   p{1.1cm}<{\centering}| 
          p{0.9cm}<{\centering}| 
 p{1.cm}<{\centering}|
 p{1.cm}<{\centering}|
 p{1.cm}<{\centering}|
 p{1.cm}<{\centering}|
 p{1.2cm}<{\centering}|
 p{1.cm}<{\centering}|
 p{1.1cm}<{\centering}|
 p{1.4cm}<{\centering}}
	
		\toprule[1pt] 
		\specialrule{0em}{0.1pt}{0.1pt}
 
\multirow{2}{*}{\#} &
\multicolumn{4}{c|}{Kernel size } & 
\multicolumn{1}{c|}{$S$ub-branch}  &
Node &
RFS & 
\multicolumn{1}{c|}{ASC} & \multicolumn{1}{c|}{AEC} & 
        \textit{pleasant} & \textit{eventful} & \textit{chaotic} & \textit{vibrant} & \textit{uneventful} & \textit{calm} & \textit{annoying} & \textit{monotonous} \\

		\cline{2-18}     
	       & 3 & 5 & 7 & 9 & \multicolumn{1}{c|}{\{Mel; Loudness\}} & $n$ & Time (s) & \textit{Acc.} (\%) & \textit{AUC} & \multicolumn{8}{c}{\textit{MSE}}
        \\
	\hline 

    1 & \CheckmarkBold  &   &   &   & \multicolumn{1}{c|}{${S_1}$: \{(3, 3); (3, 1)\}} & \multirow{4}{*}{2} & 0.76 & 93.67  &	0.913 & 0.919 &	1.071 & 	1.088 &	0.987 &	1.166 & 1.013 & 1.110 &	1.170
  \\ 
  
    2 &   & \CheckmarkBold &   &   &  \multicolumn{1}{c|}{${S_2}$: \{(5, 5); (5, 1)\}} &    &  1.44 & 93.73 & 	0.917 & 	\textbf{0.904} & 	1.056 & 1.071 & 0.980 & 	\textbf{1.141} &  \textbf{1.000} & 	\textbf{1.080} & 	1.161
 \\ 
 
    3 & &   & \CheckmarkBold &   &   \multicolumn{1}{c|}{${S_3}$: \{(7, 7); (7, 1)\}} &  & 2.12  & \textbf{94.03} & \textbf{0.921} &	 	0.910  & 	1.050 &	1.067 &	0.969 &	1.145 &	1.005 & 1.092 & \textbf{1.150}
    \\ 

    4 & &   &   &  \CheckmarkBold &  \multicolumn{1}{c|}{${S_4}$: \{(9, 9); (9, 1)\}} &  & 2.80  & 93.91 & 	0.920	& 	0.916  &	\textbf{1.049} &	\textbf{1.058} 	& \textbf{0.963}	& 1.150 &	1.006 &	1.091 &	\text{1.151}
\\ 
  
    \hline



    

    
 
    5 & \CheckmarkBold & \CheckmarkBold & \CheckmarkBold &  \CheckmarkBold &    \multicolumn{1}{c|}{${S_1} + {S_2}  + {S_3}  + {S_4}$} &  8  &  2.80 & \textbf{95.31}  &   \textbf{0.941}   &  \textbf{0.899}   &  \text{1.068} & \text{1.079}  &   \text{0.979}   & \text{1.168}  &  \textbf{0.999}  & \text{1.083}  &  \text{1.159} \\ 
    
		\specialrule{0em}{0pt}{0em}
		\bottomrule[1pt]
	\end{tabular}
	\label{tab:multi_timescale}
\end{table*}
 

 
Table \ref{tab:multi_timescale} shows the performance of SoundAQnet using features with different scales. The scale of features, i.e., the convolution receptive field size (RFS), is determined by the convolution kernel size.  
The performance of SoundAQnet with single-scale kernel branches is shown in \#1-\#4 of Table \ref{tab:multi_timescale}. 
For the convolution branch with a kernel size of 3, 5, 7, and 9, the RFS of each branch's last convolution layer relative to the input acoustic features is 76, 144, 212, and 280, respectively. The corresponding RFS in seconds is shown in Table \ref{tab:multi_timescale}. 
From \#1 to \#3, when the kernel size is increased from 3 to 7, that is, the RFS is increased from 0.76\textit{s} to 2.12\textit{s}, SoundAQnet's performance on ASC and AEC is improved, but continuing to increase the kernel size does not lead to higher accuracy. 

In Table \ref{tab:multi_timescale} \#1-\#4, the emotion-related 8D AQ regression tasks achieve good results, except for \#1.  
This indicates that the length of audio clips input to SoundAQnet needs to be greater than 0.76\textit{s} to effectively capture human-perceived AQs. 
For \#2 at the 1.44\textit{s} RFS, SoundAQnet outperforms \#1 in predicting AQs \textit{pleasant}, \textit{uneventful}, \textit{calm}, and \textit{annoying}, but a small decrease in performance is seen when increasing the kernel size further. 
For \#4 at the 2.80\textit{s} RFS, SoundAQnet outperforms options with smaller kernel sizes in predicting AQs \textit{eventful}, \textit{chaotic}, and \textit{vibrant}.
\mybluehl{The results of \#1-\#4 suggest that SoundAQnet is time-window-aware, just as people may need different time scales to perceive different AQs.}
Finally, \#5, which combines branches with different kernel sizes and RFS, performs better on ASC and AEC tasks while still performing reasonably well in most AQ regressions. \mybluehl{Other combinations of $S_1$, $S_2$, $S_3$, and $S_4$ have been explored, with more details shown in the \textbf{\textit{homepage}}.}
In short, with the cooperation of small and large-size convolution kernels, SoundAQnet extracts multiscale features suitable for the target tasks, and captures acoustic environment information from multiple perspectives, thereby improving the results.



\begin{table}[b] 
	\setlength{\abovecaptionskip}{-1mm}   
	\setlength{\belowcaptionskip}{-0mm}  
	\renewcommand\tabcolsep{1pt} 
    \vspace{-6mm}  
	\centering
	\caption{Mean performance of SoundAQnet with different methods for fusing Mel and loudness-based sub-branches on the test set.}
	\begin{tabular}{  
	p{0.2cm}<{\centering}|
	p{3.6cm}<{\centering}| 
	p{1.45cm}<{\centering}|
	p{1.6cm}<{\centering}|
    p{1.65cm}<{\centering}
 }
	
		\toprule[1pt] 
		\specialrule{0em}{0.1pt}{0.1pt}
 
\multirow{2}{*}{\#} &
\multirow{2}{*}{Fusion: (${\mathbf{M}}$, ${\mathbf{L}}$) } &   
\multicolumn{1}{c|}{ASC} & \multicolumn{1}{c|}{AEC} & AQ regression\\
		\cline{3-5}     
	    &  & \textit{Acc.} (\%) & \textit{AUC} & \textit{MSE Mean} \\
	\hline 

    1 & Addition: ${\mathbf{F}}$ = $\mathbf{M}+\mathbf{L}$ &  94.34$\pm$0.75 & 0.936$\pm$0.006 &  1.070$\pm$0.084  \\  

    2 & Concat: ${\mathbf{F}}$ = $Concat(\mathbf{M}, \mathbf{L}$) &  94.47$\pm$0.57 & 0.934$\pm$0.005 &  1.068$\pm$0.089  \\  

    3 & Hadamard: ${\mathbf{F}}$ = ${\mathbf{M}}$ $\odot$ ${\mathbf{L}}$ &  94.65$\pm$0.51 & 0.937$\pm$0.004 &  1.071$\pm$0.092\\  

    4 & Q\_Mel: ${\mathbf{F}} = Att(\mathbf{M}, \mathbf{L})$ & 88.85$\pm$2.96 & 0.865$\pm$0.008 &  1.059$\pm$0.083 \\ 

    5 & Q\_Loudness: ${\mathbf{F}} = Att(\mathbf{L}, \mathbf{M})$ &   94.54$\pm$0.99 & 0.884$\pm$0.013 &  1.040$\pm$0.082 \\ 

    6 & $Att$\_Q\_M\_Q\_L &  94.67$\pm$0.70 & 0.898$\pm$0.009 &  \textbf{1.038}$\pm$0.080  \\  

    7 & \mybluehl{CLAP attention feature fusion} &  92.61$\pm$1.05   &  0.912$\pm$0.006  & 1.066$\pm$0.080 \\ 

    8 & Graph-based &  \textbf{95.31}$\pm$0.77   &  \textbf{0.941}$\pm$0.007  & 1.054$\pm$0.091 \\
     
		\specialrule{0em}{0pt}{0em}
		\bottomrule[1pt]
	\end{tabular}
	\label{tab:SoundAQnet_fusion_method}
\end{table}

\noindent
\textit{\mybluehl{3) Ablation study on multiscale embedding fusion}}

The multiscale output is given as ${\mathbf{M}}$ = $($${m3}$, ${m5}$, ${m7}$, ${m9}$) for Mel branches and ${\mathbf{L}}$ = $($${l3}$, ${l5}$, ${l7}$, ${l9}$) for loudness branches.  
\mybluehl{We can obtain the fusion result $F$ by fusing these outputs, and then feed ${\mathbf{F}}$ into the residual embedding layer. Table \ref{tab:SoundAQnet_fusion_method} shows SoundAQnet's performance with different fusion methods. 
The mean and variance of the MSE of 8D AQ regressions are shown as an overall metric.} For \#1 in Table \ref{tab:SoundAQnet_fusion_method}, they are added; for \#2, they are concatenated; for \#3, the Hadamard product is used, where $\odot$ is the element-wise product. 
SoundAQnet performs similarly based on the fusion of \#1-\#3.
\#4-\#6 adopt the scaled dot-product attention ($Att$), a key component in the widely used Transformer \cite{Transformer}. 
\begin{equation}
\setlength{\abovedisplayskip}{1pt}
\setlength{\belowdisplayskip}{1pt} 
Attention(\mathbf{Q, K, V} )=softmax(\mathbf{QK^T} / \sqrt{d_{k} } )\mathbf{V} 
\label{self-attention}
\end{equation} 
where $\mathbf{V}$$=$$\mathbf{K}$, and $d_{k}$ is $\mathbf{K}$'s dimension. 
\mybluehl{In this case, the similarity between $\mathbf{Q}$ and $\mathbf{K}$ is exploited to adjust the information in $\mathbf{V}$, so a more informative $\mathbf{V}$ can lead to better results.}
In \#4, ${\mathbf{M}}$ acts as $\mathbf{Q}$ and ${\mathbf{L}}$ acts as $\mathbf{K}$, Mel-based representations are used as a query to adjust loudness-based representations, and the output result mainly relies on ${\mathbf{L}}$. The operation of \#5 is the opposite of \#4, and \#5 performs better than \#4 on ASC and AEC tasks. The reason is similar to the Mel-only and loudness-only models in Table \ref{tab:ablation_two_features_ASC_AEC_ISO}. Notably, both \#5 and \#6, which use loudness as $\mathbf{Q}$ to modulate Mel-based representations, perform better in AQ regressions. 
\#6, which concatenates $Att()$ in \#4 and \#5, shows the best result for AQ regressions. 
\mybluehl{\#7 uses the attention feature fusion (AFF) for variable-length audio in contrastive language-audio pretraining (CLAP) \cite{CLAP}. 
\begin{equation}
\setlength{\abovedisplayskip}{1pt}
\setlength{\belowdisplayskip}{1pt} 
X_{fusion}=\alpha X_{global} + (1-\alpha) X_{local}
\label{CLAP-aff}
\end{equation} 
where $\alpha=f_{AFF}(X_{global}, X_{local})$ is a factor obtained by AFF \cite{CLAP}.
Unlike CLAP, which only uses Mel features, SoundAQnet uses two types of features. Thus, for CLAP-AFF-based SoundAQnet, Mel-based branches and loudness-based branches calculate the corresponding $X_{fusion\_Mel}$ and $X_{fusion\_loudness}$, respectively. Then, these two are concatenated and fed into a 1-layer multilayer perceptron (MLP) for fusion.}
Overall, the graph-based multiscale embedding fusion improves ASC and AEC performance, and shows competitive overall performance in regressions of human-perceived AQs.  
\mybluehl{Source code for these fusions can be found in the \textbf{\textit{homepage}}.}

\noindent
\textit{\mybluehl{4) The impact of network depth}}

\mybluehl{
SoundAQnet contains three layers of Conv2D blocks, each referring to VGG \cite{vgg} and consisting of two convolution layers. Table \ref{tab:SoundAQnet_depth} presents the number of parameters (Param.), multiply-accumulate operations (MACs), and GPU memory required to train SoundAQnet with different depths. 
}

\vspace{-2mm} 
\begin{table}[H] 
	\setlength{\abovecaptionskip}{-1mm}   
	\setlength{\belowcaptionskip}{-0mm}  
	\renewcommand\tabcolsep{1pt} 
    \vspace{-2mm}  
	\centering
	\caption{SoundAQnet with different numbers of layers. (Batch size=32)}
	\begin{tabular}{  
	p{1cm}<{\centering}| 
p{1cm}<{\centering}| 
p{0.85cm}<{\centering}| 
p{0.85cm}<{\centering}| 
p{0.85cm}<{\centering}|     
	p{1.1cm}<{\centering}|
	p{0.8cm}<{\centering}|
    p{1.7cm}<{\centering}}
	
		\toprule[1pt] 
		\specialrule{0em}{0.1pt}{0.1pt}
 
\multirow{2}{*}{\# blocks} &   
\multirow{2}{*}{\# layers} & 
\multicolumn{1}{c|}{Para.} & 
\multicolumn{1}{c|}{MACs} & 
\multicolumn{1}{c|}{GPU} & 
\multicolumn{1}{c|}{ASC} & 
\multicolumn{1}{c|}{AEC} &
AQ regression\\
		\cline{6-8}     
	 & & (M) & (G) & (GB) & \textit{Acc.} (\%) & \textit{AUC} & \textit{MSE Mean} \\
	\hline 

    2 & 4 & 1.53 & 19.58 & 11.67 & 93.46  & 0.926  &  1.092  \\  
    3 & 6 & 2.70 & 27.68 & 13.54  &  95.31  & \textbf{0.941}  &  \textbf{1.054}  \\ 
    4 & 8 & 7.36  & 93.02 &  21.48 &  \textbf{95.60}  & 0.938  &  1.058  \\ 
    5 & 10 & 25.92 & 667.97 & 58.69  &  95.44  & 0.933  &  1.060  \\

		\specialrule{0em}{0pt}{0em}
		\bottomrule[1pt]
	\end{tabular}
	\label{tab:SoundAQnet_depth}
\end{table}

\vspace{-2mm} 
\mybluehl{
In Table \ref{tab:SoundAQnet_depth}, as SoundAQnet's depth increases, its computational overhead increases significantly but does not bring better results. SoundAQnet with the default number of blocks, three, balances model complexity, computational overhead, and achieves competitive performance in ASC, AEC, and AQ regression tasks. Due to space limitations, please also refer to the \textbf{\textit{homepage}} for code and results for different dilation rates.}

\begin{figure}[b] 
	\setlength{\abovecaptionskip}{-0mm}  
    \setlength{\belowcaptionskip}{-0cm}   
    \vspace{-6mm}
	\centerline{\includegraphics[width = 0.5 \textwidth]{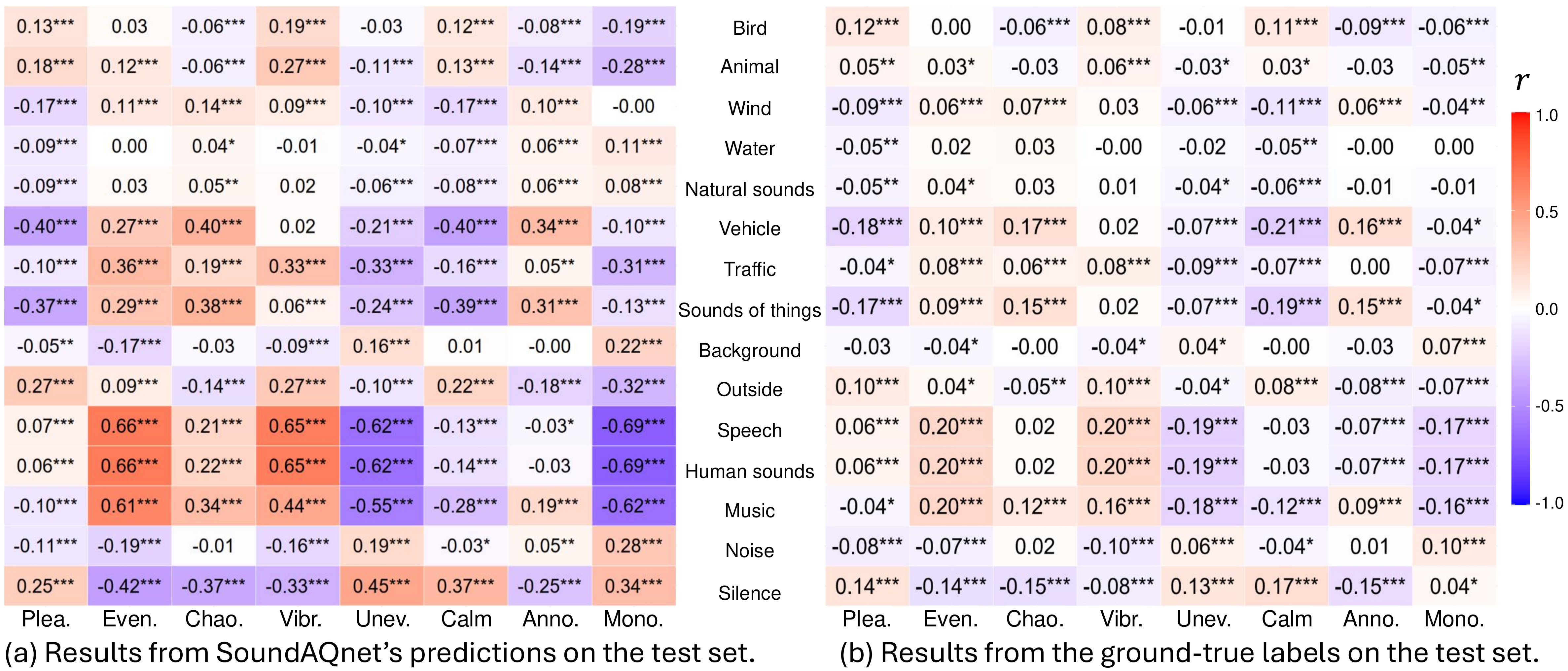}}
	\caption{Spearman's rho correlation coefficients of AE and AQ. {*}, **, and *** denote statistical significance at the 0.05, 0.01, and 0.001 levels, respectively.}
	\label{fig_event15_PAQ8}
\end{figure}

\vspace{-2mm} 
\subsection{\mybluehl{Correlation Study between AEs and AQs with SoundAQnet}}\label{RQ5} 


\mybluehl{People respond affectively and rate AQs based on recognition of various AEs, which are more related to AQ along the arousal axis \cite{schuller2012automatic}. 
This section answers the question whether SoundAQnet performs well in predicting AQs by implicitly learning the relationship between AEs and AQs.}
To investigate, Fig. \ref{fig_event15_PAQ8} (a) shows the statistical significance of the predictions given by SoundAQnet \mybluehl{on the test set of 3576 30-second binaural audio clips to analyze the relationship between AEs and the AQs they evoke}.
The Shapiro-Wilk test shows that the distributions of 15 AEs and 8D AQs are non-normal ($\alpha$ $>$ $0.05$). Thus, we use Spearman's rho for correlation analysis between AEs and AQs.  
The statistical results in Fig. \ref{fig_event15_PAQ8} (a) show that there are significant correlations between AEs and AQs.  Specifically, some AEs like \textit{`Bird'}, \textit{`Animal'}, \textit{`Outside, rural or natural' (Outside)} and \textit{`Silence'} have significant positive correlations with pleasantness and calm. In addition, some AEs like \textit{`Human sounds'}, \textit{`Music'}, and \textit{`Speech'} have significant positive correlations with eventful and vibrant, while some AEs, including  \textit{`Sound of things'} and \textit{`Vehicle'}, can significantly evoke annoyingness (Anno.) and Chaotic. This indicates SoundAQnet's capability to capture the correlation between AEs and different AQs.




To further explore how SoundAQnet learns, Fig. \ref{fig_event15_PAQ8} (b) shows the correlations on the ground-truth (GT) labels of the test set. This allows us to compare the differences in AE and AQ correlations between SoundAQnet predictions and the GT labels based on the same audio clips. Overall, the AE and AQ correlation trends in Fig. \ref{fig_event15_PAQ8} (a) and (b) are consistent. However, the correlation trend in Fig. \ref{fig_event15_PAQ8} (a) is stronger, indicating a more monotonous trend. SoundAQnet seems to accentuate correlations between specific AEs and AQs. For example, \textit{`Animal'} correlates more significantly with all 8D AQs in Fig. \ref{fig_event15_PAQ8} (a) than in (b). The stronger correlations in Fig. \ref{fig_event15_PAQ8} (a) imply that SoundAQnet favours monotonous trends by reducing noise from the relationships it identifies as important.

\begin{figure}[b] 
	\setlength{\abovecaptionskip}{-1mm}  
    \setlength{\belowcaptionskip}{-2mm}   
    \vspace{-6mm}
	\centerline{\includegraphics[width = 0.37 \textwidth]{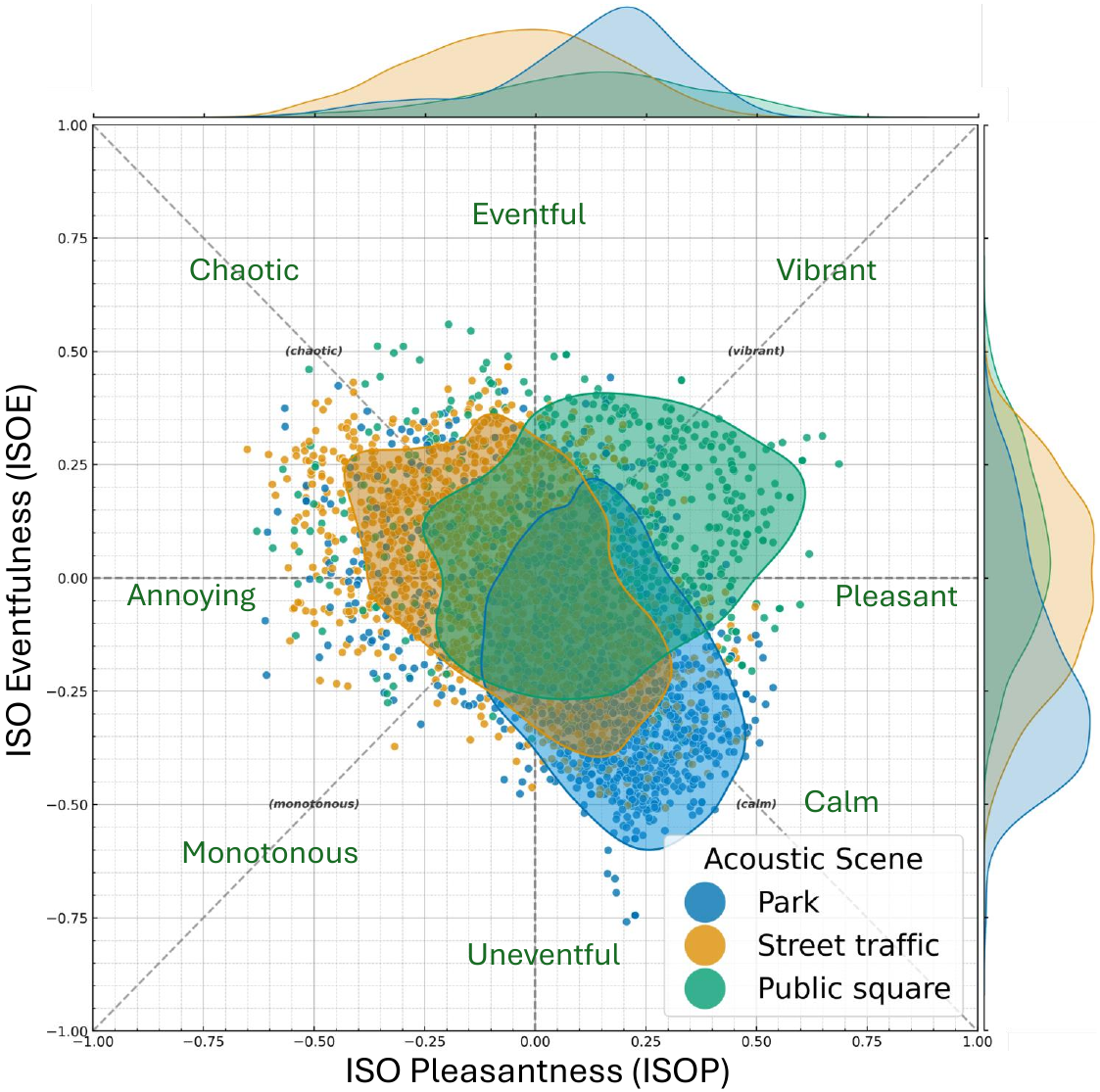}}
	\caption{The scatter plot, density distribution curves, and marginal distributions for the scenes involved are from SoundAQnet's predictions on the test set.}
	\label{fig_density_margin_plot}
\end{figure}

\vspace{-2mm} 
\subsection{\mybluehl{Analyzing ASs and AQs with SoundAQnet}}\label{RQ6}


Fig. \ref{fig_density_margin_plot} shows the scatter plots, density, and marginal distributions of audio recordings of different scenes identified by SoundAQnet on the circumflex plane of affect (AQ plane).
The density plots show that although there is a strong overlap between the distributions in the center of the plot, recordings of street traffic (ST) tend to be evaluated as more chaotic, recordings of parks tend to be rated as calmer, and recordings of public squares (PS) tend to be more vibrant. This difference between distributions is consistent with human intuitive expectations and matches previous research \cite{carvalho2024evaluating}.
The marginal distributions of ISOP suggest that people may perceive similar pleasure in the park and PS scenes; the distribution of pleasure in the park is more concentrated than in PS, and people perceive less pleasure in ST than in the park and PS. 
The marginal distributions of ISOE suggest that the PS and ST scenes have similar event richness; ST's event diversity exceeds that of PS, and people perceive less event diversity in the park than in ST. 
\mybluehl{Similar results were found based on human perception \cite{hong2015influence}. 
For more visualization plots, please see the \textbf{\textit{homepage}}.}

\vspace{-1mm} 
\subsection{\mybluehl{SoundAQnet on the Cross-cultural Soundscape Dataset}}\label{RQ7}



\mybluehl{The soundscape standard ISO/TS 12913-2 \cite{ISO_soundscape_data} gives researchers some freedom to incorporate cultural differences, but the basic idea behind the PAQ in \cite{ISO_soundscape_data} is still to try to remove cultural bias, starting with language. 
Soundscape attributes translation project (SATP) \cite{SATP} aims at removing bias due to language and culture. The SATP dataset is a cross-cultural, multi-lingual, international, and cross-institutional dataset containing recordings with ISO-standard AQ labels. 
The SATP contains 19089 samples in 18 languages, consisting of 707 participants from 29 different national institutions evaluating 27 30-second binaural audio recordings according to ISO/TS 12913-2:2018 \cite{ISO_soundscape_data}, in the institution’s own language in each region. There are an average of 32 participants per institution.}

\mybluehl{To evaluate SoundAQnet's generalization ability across cultures, first, the Wilcoxon signed-rank test is used to compare the model's predictions on SATP with the ratings of 707 participants. The results show no significant difference between SoundAQnet's predictions and the average ratings of 707 participants ($p>0.05$); that is, SoundAQnet performs similarly to the participants.
Then, to explore the model performance in specific cultures, the consistency analysis of two cases in Europe and Asia is performed using Intraclass Correlation Coefficient (ICC)~\cite{shrout1979intraclass} analysis. As shown in Table \ref{tab:SATP_UCL_NTU}, SoundAQnet's predictions show significant consistency with participants at University College London (UCL) on all AQs ($p<0.05$); and with participants at Nanyang Technological University (NTU) on 7 AQs ($p<0.05$) except Calm ($p>0.05$). This indicates that SoundAQnet shows significant consistency on multiple affective dimensions across two different cultures, including Asian and Western cultures.}


\vspace{-3mm}
\begin{table}[H] 
	\setlength{\abovecaptionskip}{-1mm}   
	\setlength{\belowcaptionskip}{-2mm}  
	\renewcommand\tabcolsep{1pt} 
	\centering
	\caption{ICC analysis for two different cultures (* is $p<0.05$); ICC ranges from 0 (no consistency) to 1 (perfect consistency).
    }
	\begin{tabular}{  
	p{1.5cm}<{\centering}|
	p{0.8cm}<{\centering}|
 p{0.8cm}<{\centering}|
	p{0.8cm}<{\centering}|
	p{0.8cm}<{\centering}|
 p{0.8cm}<{\centering}|
 p{0.8cm}<{\centering}|
 p{0.8cm}<{\centering}|
 p{0.8cm}<{\centering}}
	
		\toprule[1pt] 
		\specialrule{0em}{0.1pt}{0.1pt}

	      PAQ & Plea & Even & Chao & Vibr & Unev & Calm & Anno & Mono\\
	\hline 

    UCL\_ICC &  0.89* & 0.85*  &  0.89*   &  0.62*  & 0.82* & 0.87* & 0.82*& 0.57*\\  
    \hline
    NTU\_ICC & 0.78*  & 0.60*  &  0.90*   &  0.65*  & 0.61* &  0.00  & 0.86* &  0.53* \\
  
		\specialrule{0em}{0pt}{0em}
		\bottomrule[1pt]
	\end{tabular}
	\label{tab:SATP_UCL_NTU}
\end{table}

\vspace{-2mm}
\mybluehl{The results on SATP dataset show that SoundAQnet captures some degree of intra-cultural consistency. Prior studies show that LLMs generate different responses when queried in different languages, each embedding elements of its local culture \cite{buyl2024large}.
Integrating SoundAQnet with LLMs may provide a good start for future studies aimed at fine-tuning the model to specific cultural contexts. In addition, we acknowledge that cultural differences in soundscapes are complex. Affective responses to soundscapes are inherently subjective and affected by personal, linguistic, and cultural factors, which is also one of the key challenges in soundscape studies.
}

\section{\mybluehl{Human evaluation (SoundSCaper experiment)}} \label{section_human_Evaluation}

To assess the quality of captions generated by the SoundSCaper system, crowdsourced human evaluation is used to compare captions from SoundSCaper with captions annotated by two soundscape experts after cross-checking each other.

\vspace{-3mm}
\subsection{Experimental Design for Quality Assessment}
\label{section_human_Evaluation_design}   

The study employs a within-subjects design to evaluate soundscape captions from SoundSCaper and human experts. The sample size calculation is performed using G*Power \cite{Gpower}. The results of the calculation indicated that a sample size of 30 audio samples with $\alpha=0.05$ and an assumed Effect Size of 0.5 for the Wilcoxon signed rank test achieved the pre-statistical power of 83.3\%.
Thus, the evaluation contains 60 audio clips from two distinct datasets. Dataset 1 (D1) contains 30 randomly selected samples with the same sound pressure levels (SPLs) from this paper's test set. 
\mybluehl{
Dataset 2 (D2) has 30 samples randomly selected from 5 external, i.e., model-never-seen, audio scene datasets, which are DCASE 2018/2019 \cite{dcase2019_t1a}, ISD \cite{ISD}, LITIS-Rouen \cite{rouen} and a road sound dataset \cite{coensel2011effects} of freeways with extreme noise environments commonly seen in daily life.
}
The training set used in this paper contains 3 types of acoustic scenes, so recordings related to the 3 AS labels are selected from the 5 external datasets. Finally, the total duration of the D2 candidate data pool is about 1177 hours. 
The audio clips in D2 vary from 10 to 30 seconds with various SPLs without any limitations. Hence, D2 is used mainly to test the generalization performance of the SoundSCaper system.

\subsubsection{Soundscape expert annotations} Two soundscape experts listen to randomly ordered samples and write captions in a style similar to the SoundSCaper caption example. This is done to ensure the consistency of caption styles generated by SoundSCaper and experts to prevent bias caused by participants guessing the caption's origin based on different styles.

\subsubsection{SoundSCaper captions} 
As described in Section \ref{sec::LLM}, SoundSCaper automatically generates target descriptions.

Finally, 120 soundscape captions are evaluated, 60 of which are derived from the proposed SoundSCaper system, and the remaining 60 are annotated by the two experienced soundscape experts. These captions are randomized. 
\mybluehl{These captions were evaluated by a jury of 16 audio/soundscape experts and another jury of 16 laypersons, totaling 32 participants from 8 countries, including the UK, Singapore, Belgium, Canada, and France.}
Human assessment instructions, assessment data, and statistics' metadata are publicly available on the \textbf{\textit{homepage}}.
\mybluehl{Based on a self-assessment of the study's risks, ethical approval for this research was obtained from the Faculty of Engineering and Architecture of Ghent University.}

 

\vspace{-3mm}
\subsection{Soundscape Caption Evaluation Metrics}

Inspired by \cite{kasai2021transparent}, we introduce the Transparent Human Benchmark for Soundscapes (THumBS) as a metric for the ASSC task. This indicator consists of precision, recall, and three other types of penalty items targeting specific defects.
 
\subsubsection{Precision and recall $\in [1, 5]$}
 
Precision (\textit{P}) measures the accuracy of captions in describing the soundscape, specifically how well the caption's details match the actual sounds.
Recall (\textit{R}) evaluates the extent to which the caption captures the comprehensive range of salient information (e.g., objects, attributes, relations) present in the soundscape.

 
\subsubsection{Penalty items $\in [-2, 0]$}
 
Fluency (\textit{F}) assesses captions' textual quality as English prose, independent of its content accuracy.  
Conciseness (\textit{C}) is used for repetitive descriptions.
Irrelevance (\textit{I}) is applied to the captions with details not present in the soundscape or unrelated to the sound content.

\subsubsection{THumBS score}
The final score can be  calculated as
\begin{equation}
\setlength{\abovedisplayskip}{1pt}
\setlength{\belowdisplayskip}{1pt} 
Score = (P + R) / 2 + F + C + I
\label{human_score}
\end{equation}

\mybluehl{Due to limited space, we fully explain these metrics in the participant instructions presented on the \textbf{\textit{homepage}}.}

\begin{figure*}[b] 
	\setlength{\abovecaptionskip}{-2mm}  
    \setlength{\belowcaptionskip}{0cm}   
    \vspace{-5mm}  
	\centerline{\includegraphics[width = 0.82 \textwidth]{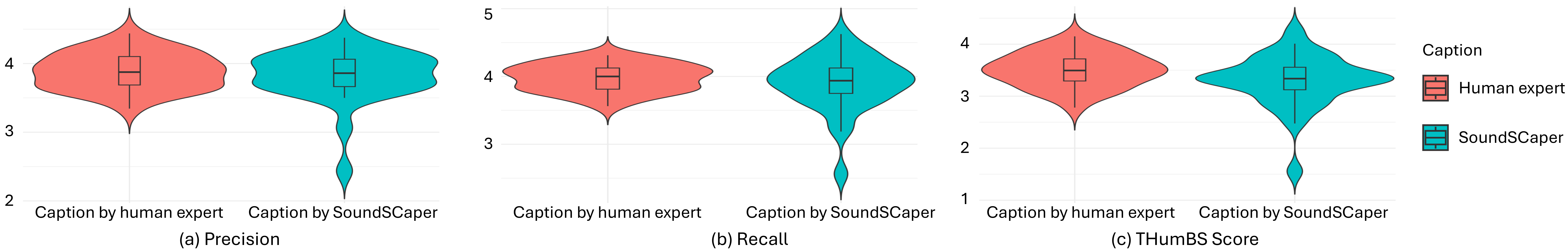}}
	\caption{\textit{P}, \textit{R} and \textit{THumBS} score of soundscape captions given by a jury of 16 audio/soundscape experts on the dataset D1.}
	\label{Flac_3_box_legend}
\end{figure*}

\begin{figure*}[b] 
	\setlength{\abovecaptionskip}{-2mm}  
    \setlength{\belowcaptionskip}{0cm}  
    \vspace{-4mm}  
	\centerline{\includegraphics[width = 0.82 \textwidth]{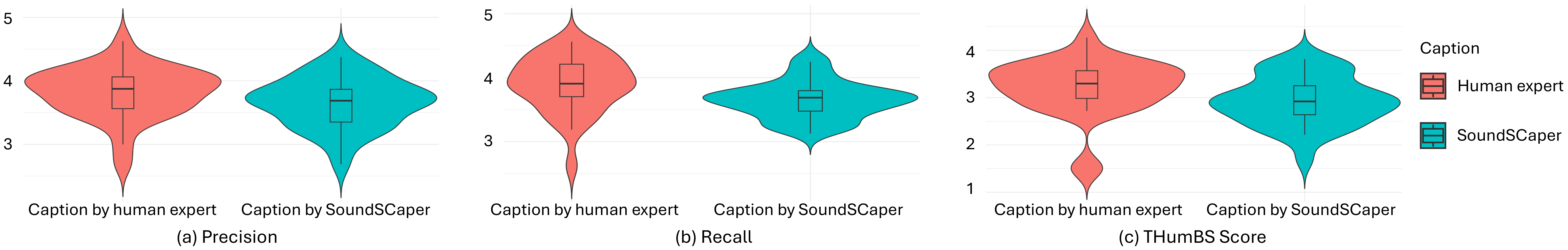}}
	\caption{\textit{P}, \textit{R} and \textit{THumBS} score of soundscape captions given by a jury of 16 audio/soundscape experts on the external dataset D2.}
	\label{WAV_3_box_legend}
\end{figure*}

\vspace{-2mm}
\subsection{\mybluehl{Professionals' Evaluations}}\label{case}


\noindent
\textit{\mybluehl{1) Comparison of SoundSCaper and expert captions}}


\vspace{-4mm}
\begin{table}[H]\footnotesize
	\setlength{\abovecaptionskip}{-1mm}   
	\setlength{\belowcaptionskip}{-0.2cm}  
	\renewcommand\tabcolsep{1pt} 
	\centering
	\caption{Comparison of soundscape caption quality from soundscape expert (E) and SoundSCaper (S) on datasets D1 and D2.}
	\begin{tabular}{  
	p{0.2cm}<{\centering}|
	p{0.3cm}<{\centering}|
        p{1.2cm}<{\centering}|
	p{1.2cm}<{\centering}|
	p{1.35cm}<{\centering}|
 	p{1.4cm}<{\centering}|
   	p{1.35cm}<{\centering}|
 p{1.2cm}<{\centering} }
	
		\toprule[1pt] 
		\specialrule{0em}{0.1pt}{0.1pt} 
\hline 

    D &   & \textit{precision} &  \textit{recall}  &  \textit{fluency} & \textit{conciseness} & \textit{irrelevance} & Score\\
    \hline  

    \multirow{2}{*}{1} &   E  & \textbf{3.84}$\pm$0.30 &  \textbf{3.93}$\pm$0.21  &  \textbf{-0.10}$\pm$0.07 & \textbf{-0.14}$\pm$0.12 & -0.22$\pm$0.12 & \textbf{3.43}$\pm$0.35 \\  
     & S  & 3.79$\pm$0.39 &   3.86$\pm$0.43  & -0.12$\pm$0.09  & -0.30$\pm$0.15 & \textbf{-0.18}$\pm$0.15 &3.22$\pm$0.53 \\   
         \hline  
      \multirow{2}{*}{2}  &   E  & \textbf{3.79}$\pm$0.39 &  \textbf{3.88}$\pm$0.43  &  \textbf{-0.15}$\pm$0.11 & \textbf{-0.26}$\pm$0.12& \textbf{-0.26}$\pm$0.19 & \textbf{3.16}$\pm$0.58\\  
    & S   & 3.64$\pm$0.39 &   3.64$\pm$0.29  & -0.16$\pm$0.10  & -0.27$\pm$0.16& -0.29$\pm$0.14&  2.91$\pm$0.52\\  
		\specialrule{0em}{0pt}{0em}
		\bottomrule[1pt]
	\end{tabular}
	\label{tab:DATASET1_DATASET2_EVALUATION}
\end{table}

\vspace{-3mm}
In the within-subject design study, the Shapiro-Wilk normality (SWN) test result shows that precision, recall and the final score data do not follow a normal distribution. Hence, we use the non-parametric Wilcoxon signed-rank (WSR) test. The results in Table \ref{tab:DATASET1_DATASET2_EVALUATION} show that there is no significant difference between captions generated by SoundSCaper and those offered by soundscape experts on the final score ($p=0.128$), and no significant difference between the two in terms of precision ($p$ = $0.34$) and recall ($p$ = $0.44$). This means that the quality of soundscape captions generated by SoundSCaper is comparable to that of soundscape expert-annotated captions. Fig.~\ref{Flac_3_box_legend} details the precision, recall and final THumBS score in the evaluation of dataset D1. The horizontal line bisecting the box is the median, which coincides with the top line; the red dot represents the mean. The top and bottom borders of the box represent the 25th and 75th percentiles, respectively.


The intraclass correlation coefficient (ICC)~\cite{shrout1979intraclass} is used to assess the reliability of 16 raters' average ratings for 120 captions across 60 audio samples for Precision, Recall, Fluency, Conciseness, and Irrelevance. 
ICC values ($0.345\sim0.551$) show moderate to high agreement, with significant consistency ($p < 0.001$) and upper bounds of the confidence intervals above 0.6 for most criteria. These results indicate that 16 raters are sufficient to provide reliable assessments in this study.

  

\mybluehl{For the mixed external dataset D2, 
the SWN test results indicate that the final score data on D2 do not follow a normal distribution ($p<0.05$).} 
Hence, we use a non-parametric WSR test. Table~\ref{tab:DATASET1_DATASET2_EVALUATION} shows that expert-annotated captions scored slightly higher than the captions generated by SoundSCaper; however, the WSR test shows that there is no significant difference between the two in the final scores ($p = 0.051$). As $p$ is close to the significant level, we evaluate the ratings on the precision, recall, and penalty items, including fluency, conciseness, and irrelevance, separately.  
The distributions in Fig.~\ref{WAV_3_box_legend} show that the precision and recall ratings follow a normal distribution, while the penalty items do not. Therefore, we use the paired t-test for precision and recall ratings; the result implies that there is no significant difference between the SoundSCaper-generated and expert-annotated captions on precision rating ($p$ = $0.19$) while there is a significant difference in recall rating ($p$ = $0.028$), which is not surprising as SoundAQnet is not trained on those datasets and the AE labels are also limited. The WSR test implies that there is no significant difference between SoundSCaper-generated and expert-annotated captions on fluency ($p$ = $0.33$), conciseness ($p$ = $0.97$), and irrelevance ($p$ = $0.21$). In summary, SoundSCaper has good generalization performance and adaptability, even though the recall rating of SoundSCaper-generated captions is significantly lower than that of the expert-annotated captions, and a competitive final score is still achieved.

\noindent
\textit{\mybluehl{2) Case study on the differences}}


The violin plot in Fig.~\ref{Flac_3_box_legend} (c) reveals that the score distribution of SoundSCaper shows a slightly lower tail compared to that of the soundscape experts, indicating that SoundSCaper underperforms on some audio clips. Here, we explore the largest gap in final scores between SoundSCaper and human experts by subtracting the final score of SoundSCaper captions from that of soundscape expert annotations. The maximum value in the difference sequence is 1.99, observed in the sample \textit{``28.flac"}. The soundscape captions are:



\vspace{-0.5mm}
\begin{figure}[H] \setlength{\abovecaptionskip}{0mm} 
    \setlength{\belowcaptionskip}{0cm}  
    \vspace{-2mm}  
	\centerline{\includegraphics[width = 0.5 \textwidth]{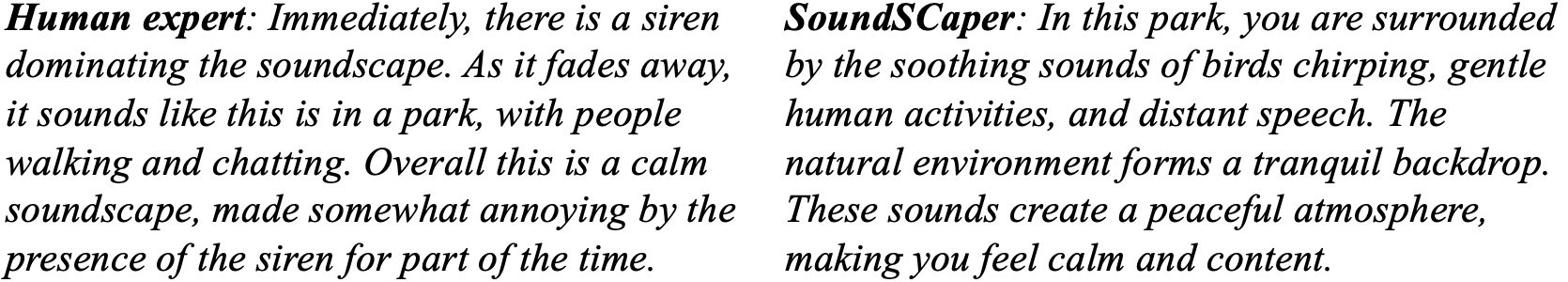}}
	\label{case1}
\end{figure}

\vspace{-3mm}
\mybluehl{In this case, the expert outperforms SoundSCaper, and emphasizes the disruptive presence of a siren in the park scene, highlighting its significant impact on the soundscape's calmness. Conversely, SoundSCaper paints a serene picture, not mentioning sirens and focusing only on peaceful elements such as birds chirping and distant speech. The reason is that there are no sirens in the 15 classes of AE labels in the dataset used for training the SoundAQnet, and hence it does not recognize this sound.}

Next, we explore aspects where SoundSCaper outperforms human experts by subtracting the final score of expert-annotated captions from that of SoundSCaper captions. The maximum score difference is 0.84, the sample is \textit{``26.flac"}, and its corresponding soundscape captions are:



\begin{figure}[H] 
	\setlength{\abovecaptionskip}{0mm} 
    \setlength{\belowcaptionskip}{0cm}  
    \vspace{-2mm}  
	\centerline{\includegraphics[width = 0.5 \textwidth]{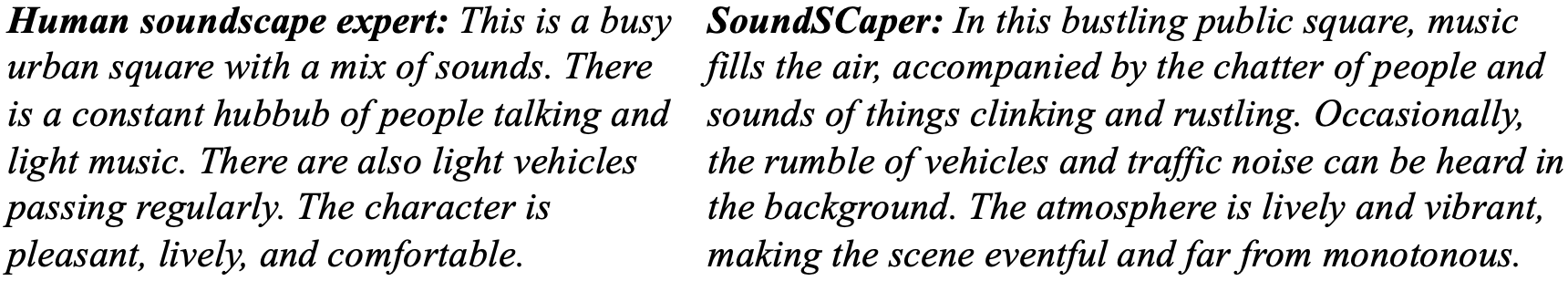}}
	\label{case1}
\end{figure}

\vspace{-3mm}
In this case, the SoundSCaper caption is more appreciated by professionals. The expert's caption captures a mixture of pleasant, comfortable, and lively sounds in an urban square. SoundSCaper predominantly depicts the scene's lively and vibrant aspects, such as music and people's chatter, while downplaying vehicle noise. 
One possible explanation for this difference is that emotional feelings are subjective. 
\mybluehl{People from various experiences and socio-cultural backgrounds may feel the same sound differently.} 
Compared to descriptions with experts' individual responses to AQs, the AQ values predicted by SoundAQnet, \mybluehl{trained on the ARAUS dataset of 25248 samples assessed by 605 participants, may be more consistent, less personalized, and more acceptable to other participants.}

\vspace{-4mm}
\subsection{\mybluehl{Layperson Evaluations}}


\mybluehl{We employ the within-subject design for the layperson evaluation experiment. The SWN test shows that the final score data for D1 do not follow a normal distribution. Hence, we use the non-parametric WSR test. 
Table \ref{tab:DATASET1_DATASET2_EVALUATION_enduser} shows that there is no significant difference between captions generated by SoundSCaper and those provided by soundscape experts in terms of the final score ($p=0.136$) by the layperson jury. That is, in the opinion of laypersons, the quality of captions generated by SoundSCaper is comparable to that of soundscape expert-annotated captions.}


\mybluehl{For the external dataset D2, the SWN test results show that the final score data do not follow a normal distribution, so the non-parametric WSR test is used.
Table \ref{tab:DATASET1_DATASET2_EVALUATION_enduser} shows that in the evaluation by a layperson jury, SoundSCaper captions scored 3.24, slightly higher than the expert-annotated captions. However, the WSR test results show no significant difference in the final score between the two ($p = 0.90$).}

\vspace{-4mm}
\begin{table}[H]\footnotesize
	\setlength{\abovecaptionskip}{-1mm}   
	\setlength{\belowcaptionskip}{-0.2cm}  
	\renewcommand\tabcolsep{1pt} 
	\centering
	\caption{\mybluehl{Layperson evaluation of soundscape caption from soundscape expert (E) and SoundSCaper (S) on datasets D1 and D2.}}
	\begin{tabular}{  
	p{0.2cm}<{\centering}|
	p{0.3cm}<{\centering}|
        p{1.2cm}<{\centering}|
	p{1.2cm}<{\centering}|
	p{1.35cm}<{\centering}|
 	p{1.4cm}<{\centering}|
   	p{1.35cm}<{\centering}|
 p{1.2cm}<{\centering} }
	
		\toprule[1pt] 
		\specialrule{0em}{0.1pt}{0.1pt} 
\hline 

    D &   & \textit{precision} &  \textit{recall}  &  \textit{fluency} & \textit{conciseness} & \textit{irrelevance} & Score\\
    \hline  

    \multirow{2}{*}{1} &   E  & 3.86$\pm$0.19 &  3.82$\pm$0.21  &  \textbf{-0.13}$\pm$0.08 & \textbf{-0.10}$\pm$0.09 & -0.18$\pm$0.12 & \textbf{3.43}$\pm$0.33 \\  
     & S  & \textbf{3.90}$\pm$0.25 &   \textbf{3.84}$\pm$0.27  & -0.15$\pm$0.10  & -0.21$\pm$0.13 & -0.18$\pm$0.10 &3.33$\pm$0.36 \\   
         \hline  
      \multirow{2}{*}{2}  &   E  & \textbf{3.75}$\pm$0.38 &  \textbf{3.75}$\pm$0.35  &  -0.16$\pm$0.09 & -0.18$\pm$0.12& -0.21$\pm$0.18 & 3.20$\pm$0.58\\  
    & S   & 3.72$\pm$0.40 &   3.67$\pm$0.36  & \textbf{-0.12}$\pm$0.11  & \textbf{-0.17}$\pm$0.15& \textbf{-0.17}$\pm$0.12&  \textbf{3.24}$\pm$0.52\\  
		\specialrule{0em}{0pt}{0em}
		\bottomrule[1pt]
	\end{tabular}
	\label{tab:DATASET1_DATASET2_EVALUATION_enduser}
\end{table}


\vspace{-3mm}
\mybluehl{In addition, we compare the final scores given by 16 experts and 16 laypersons for both expert-annotated and SoundSCaper-generated captions. We use the paired t-test for expert ratings on both D1 and D2, as they follow the normal distribution. As for the final scores for SoundSCaper-generated captions, we employed the WSR test for both D1 and D2, as their ratings do not follow a normal distribution. The paired t-test results show no significant differences between the ratings of 16 experts and 16 laypersons for the expert-annotated captions on both D1 ($p = 0.32$) and D2 ($p = 0.54$). 
As for the final scores for SoundSCaper-generated captions, the WSR results show that there is no significant difference between the expert ratings and laypersons' ratings ($p = 0.58$) on D1; however, the laypersons' ratings for SoundSCaper-generated captions are significantly higher than the expert ratings on D2 ($p<0.01$), laypersons rate the SoundSCaper-generated captions slightly higher than the expert annotations. This implies that laypersons are less sensitive to inconsistencies or subtle errors in soundscape captions compared to the experts, and appreciate the SoundSCaper-generated captions more.}


\vspace{-2mm}
\subsection{\mybluehl{Comparison with AAC systems}}

\vspace{-4mm}
\begin{table}[H] 
	\setlength{\abovecaptionskip}{-1mm}   
	\setlength{\belowcaptionskip}{-0.2cm}  
	\renewcommand\tabcolsep{1pt} 
	\centering
	\caption{\mybluehl{Average results of AAC systems and SoundScaper on various NLP metrics of the ASSC task on datasets D1 and D2, using soundscape expert captions as ground-truth sentences.}}
	\begin{tabular}{  
	p{0.2cm}<{\centering}|
	p{2cm}<{\centering}| 
	p{0.9cm}<{\centering}| 
	p{0.9cm}<{\centering}| 
    p{1.2cm}<{\centering}| 
    p{1.1cm}<{\centering}| 
	p{1.2cm}<{\centering}| 
	p{0.9cm}<{\centering}
 } 
		\toprule[1pt] 
		\specialrule{0em}{0.1pt}{0.1pt}
 
\multirow{2}{*}{\#} &
\multirow{2}{*}{System}&
\multirow{2}{*}{BLEU} &
\multicolumn{3}{c|}{ ROUGE-L} &   
\multirow{2}{*}{METEOR}& 
\multirow{2}{*}{CIDEr}\\
		\cline{4-6}     
   & & & Recall & Precision & F1 Score & & \\
	\hline 

    1 & P-LocalAFT & 0.0249	 & 0.1001 & 	\textbf{0.3428} & 	0.1537	 & 0.0684	 & 0.0029 \\

    2 & ConvNeXt-Trans & 0.1115 & 	0.1454	 & 0.2542 & 	0.1831 & 	0.1168 & 	0.0917   \\

    3 & GAMA & 0.0879 & 	0.1280 & 	 0.2852	 & 0.1694	 & 0.1009	 & 0.0849   \\
    
    4 & Qwen-Audio & 0.1032	 & 0.1479	 & 0.2729 & 	0.1868 & 	0.1082	 & 0.1213   \\

    5 & SoundSCaper & \textbf{0.1901}	 & \textbf{0.2277}	 & 0.2131 & 	\textbf{0.2150}	 & \textbf{0.1610}	 & \textbf{0.2745}  \\
    
    \specialrule{0em}{0pt}{0em}
		\bottomrule[1pt]
	\end{tabular}
	\label{tab:NLP_metric}
\end{table}

\vspace{-2mm}
\mybluehl{In this section, SoundSCaper is compared with AAC systems with and without LLMs. Table \ref{tab:NLP_metric} shows the performance of typical sequence-to-sequence (S2S) systems without LLM trained on the AudioCaps and Clotho-v2 datasets: P-LocalAFT \cite{xiao2022local} and ConvNeXt-Trans \cite{labb2024conette}, and audio-LLMs such as GAMA \cite{ghosh2024gama} and Qwen-Audio \cite{chu2023qwen}. The audio encoder of ConvNeXt-Trans is ConvNeXt pre-trained on AudioSet \cite{audioset}, and the decoder consists of a Transformer decoder with a structure similar to GPT. Given the excellent performance of ConvNeXt-Trans on AAC tasks \cite{labb2024conette}, we fine-tune ConvNeXt-Trans on the training set of this paper as the baseline system for soundscape audio-to-text with affective qualities. 
GAMA \cite{ghosh2024gama} uses the audio spectrum transformer (AST), which consists of 12 Transformer encoder layers, for audio encoding, and uses multiple models such as multi-layer aggregators and Q-Former for enhancement, and then uses GPT-4 for caption generation. The number of parameters and the number of multiply-add-accumulate operations (MACs) required for AST in GAMA are 73.61MB and 108.99 G, respectively, while the corresponding SoundAQnet in SoundSCaper is only 2.70MB and 27.68 G. In terms of both the number of parameters and the model complexity, the proposed acoustic model, SoundAQnet in SoundSCaper, is much smaller than AST in GAMA. In short, Table \ref{tab:NLP_metric} shows that SoundSCaper performs better than the LLM-free S2S AAC systems and the audio-LLMs on several NLP metrics. 
Please note that the soundscape expert captions have no influence on SoundSCaper's design, implementation, and training. In fact, they are collected after SoundSCaper is completed.}

\mybluehl{For source codes and data to fine-tune the audio-to-text baseline ConvNeXt-Trans for soundscapes, as well as scripts for NLP metrics, please see the \textbf{\textit{homepage}}.}

\subsection{\mybluehl{Further Discussions}}  

\vspace{-1mm}
\mybluehl{In the case study in Section \ref{case}, soundscape experts provided more specific and context-aware soundscape captions; e.g., they pointed out that the siren dominates the soundscape first, and after it fades away, people are chatting, which increased the spatiality and realism of the soundscape. 
In contrast, SoundSCaper mentions dominant AEs but lacks information about the corresponding sound sources and their spatial and temporal distribution. Due to the limited AS and AE labels in the used dataset, SoundAQnet cannot capture subtle but key sounds (such as short sirens) on the case \textit{``28.flac"} as soundscape experts, resulting in descriptions lacking details. In addition, individuals from different cultural backgrounds have different AQs for soundscapes. The descriptions of the two soundscape experts from the UK and Sweden are based on personal experience, perceptions, and specialized training, and have a certain subjective style. Compared with the individual AQ responses of soundscape experts, the AQ values predicted by SoundAQnet, which is trained based on 25248 samples assessed by 605 participants, are more agreeable to other assessors in the case \textit{``26.flac"}. This is consistent with SoundAQnet's cross-cultural performance on the SATP dataset, which contains 19089 samples in 18 languages and comprises 707 participants from 29 national institutions as discussed in Section \ref{RQ7}. In short, SoundSCaper's acoustic model, SoundAQnet, can effectively predict soundscape AQ in human evaluation based on datasets D1 and external mixed D2 and shows certain generalization abilities on the SATP dataset across countries and cultures. However, SoundSCaper still faces challenges posed by the inherent bias and subjectivity of individuals' affective response to sound and socio-cultural differences, which are key issues in soundscape research.}

\mybluehl{To assess the quality of ASSC, we investigated whether people would rate SoundScaper-generated and human-annotated captions differently after listening to soundscape recordings. 
In evaluations by 16 experts and 16 laypersons, captions produced by SoundSCaper are rated similarly to those produced as a consensus of two experts from different countries. 
The evaluation results show no significant difference in the rating behavior between the 16 experts and 16 laypersons. 
However, the 16 laypersons rated SoundSCaper higher on the D2 dataset, probably because laypersons are not as sensitive as experts to subtle differences and imprecision in soundscape captions.}

\mybluehl{In contrast to many AAC systems, SoundSCaper is not trained on datasets with captions. However, in Table \ref{tab:NLP_metric}, where captions annotated by soundscape experts are used as ground truth, SoundSCaper outperforms AAC systems with and without LLM in NLP metrics. This shows that SoundSCaper outperforms general AAC systems in the ASSC task. However, since this study is the first attempt at ASSC, it faces some limitations: (1) the lack of datasets, preferably audiovisual datasets containing 360-degree videos and spatial audio, with high-quality soundscape captions produced by professionals; (2) the focus on common outdoor soundscapes, although the ARAUS dataset is based on USotW that contains recordings from 12 cities in 3 continents, it does not include extreme and rare soundscapes; (3) the subjectivity of individual perception of soundscapes and the differences in various regions and cultures need to be further explored. The proposed SoundSCaper's generalization performance in practical applications still needs to be optimized with more diverse data to ensure its adaptability to a wider range of soundscapes and user groups.}

\vspace{-2mm}
\section{Conclusion}\label{section_conclusion}

\mybluehl{Describing a soundscape, the acoustic environment as it is perceived and understood by people in context, is a cumbersome task. 
This paper formalizes the affective soundscape captioning (ASSC) task and designs the SoundSCaper system for this purpose. 
SoundSCaper used a novel lightweight acoustic model, SoundAQnet, to classify ASs and AEs and to predict AQs from soundscape, and also customized an LLM with prompt engineering techniques to turn such information into textual descriptions.  
Our work is significantly different from previous audio captioning studies, because it is the first attempt to jointly model the AS and AE in acoustic environments and the corresponding human-perceived AQ in soundscapes, and also the first attempt to connect the classification of AS and AE in acoustic environments with human affective descriptions of soundscapes.
Comprehensive experiments and human evaluations have been performed to demonstrate the effectiveness of the proposed system for ASSC, as compared to relevant baselines. Overall, SoundSCaper offers competitive performance in human subjective evaluation and various objective captioning metrics, and the generated captions are comparable to those annotated by soundscape experts.}


\mybluehl{
Given that this study is the first attempt at ASSC, it faces several limitations, such as the lack of available datasets with fully and carefully annotated soundscape recordings, and limited types of soundscapes and acoustic environments involved. 
In addition, AASC faces challenges posed by the inherent bias and subjectivity of individuals in their affective responses and socio-cultural differences, which are key issues in soundscape studies. Hence, the generalization of SoundSCaper needs to be further improved through diversified datasets and cross-scene and cultural validation. 
Introducing diverse large-scale data also helps the LLM in SoundSCaper fit AQs, explore scaling laws in AASC, and reduce the chance of LLM hallucinations, which remains a common issue.
In addition, SoundAQnet performs comparably with human participants on the cross-country SATP dataset, implying that it has the potential to serve in unseen cross-cultural soundscape AQ assessments.
}

\vspace{-2mm}
\ifCLASSOPTIONcompsoc
  \section*{Acknowledgments}
\else
  \section*{Acknowledgment}
\fi





\mybluehl{We appreciate the associate editor and all five reviewers for their insightful and helpful comments.}
We appreciate Dr. Francesco Aletta for valuable discussions and Dr. Gunnar Cerwen for professional soundscape captions. 
\mybluehl{
We thank 16 soundscape experts/professors in the human assessment, and another 16 general users. Due to limited space, we list their names on \textbf{\textit{homepage}} to express our deep gratitude.
}






\ifCLASSOPTIONcaptionsoff
  \newpage
\fi

\bibliographystyle{IEEEbib}
\vspace{-2mm}
\bibliography{main}

\end{document}